\documentclass[11pt]{article}
\usepackage[a4paper,bottom=4.2cm,top=1cm,head=3cm,width=18cm,dvipdfm]{geometry}
\evensidemargin=\oddsidemargin

\usepackage[linktocpage=true,bookmarks=true,bookmarksnumbered=true,breaklinks=true,pdfpagemode=Fullscreen,pdfstartview=FitBH]{hyperref}

\usepackage[dotinlabels]{titletoc}
\usepackage{titlesec}

\usepackage{graphicx}
\usepackage{dcolumn}
\usepackage{bm}
\usepackage{mathrsfs}
\usepackage{epsfig}
\usepackage{verbatim}
\usepackage{amsmath,amssymb}
\usepackage[dvips]{color}
\usepackage{hhline}
\usepackage{srcltx}
\usepackage{multirow} 
\usepackage{amsfonts}
\usepackage{bm}
\usepackage{feynmp}
\usepackage{extarrows}
\usepackage{slashed}
\usepackage{subfigure}
\allowdisplaybreaks[4]

\renewcommand{\thefootnote}{\fnsymbol{footnote}}
\newcommand{\beq}{\begin{equation}}
\newcommand{\eeq}{\end{equation}}
\newcommand{\bq}{\begin{equation}}
\newcommand{\eq}{\end{equation}}
\newcommand{\ba}{\begin{array}}
\newcommand{\ea}{\end{array}}
\newcommand{\beqa}{\begin{eqnarray}}
\newcommand{\eeqa}{\end{eqnarray}}
\newcommand{\beqs}{\begin{subequations}}
\newcommand{\eeqs}{\end{subequations}}

\def\nn{\nonumber}

\def\dis{\displaystyle}

\def\({\left(}
\def\){\right)}

\def\hf{\frac{1}{2}}

\def\End{\end{document}}
\newcommand{\FR}[2]{\displaystyle\frac{\,{#1}\,}{#2}}
\newcommand{\fr}[2]{\mbox{$\frac{\,{#1}\,}{#2}$}}
\newcommand{\n}{\nonumber}
\renewcommand{\rm}{\mathrm}
\def\bge{\begin{equation}}
\def\ede{\end{equation}}
\def\bga{\begin{aligned}}
\def\eda{\end{aligned}}
\newcommand{\order}[1]{\mathcal{O}({#1})}

\def\ga{\gamma}

\def\de{\delta}

\def\De{\Delta}

\def\MP{M_{\textrm{Pl}}^{}}

\def\leqq{\leqslant}
\def\geqq{\geqslant}

\def\hf{\frac{1}{2}}

\def\dif{\partial}

\def\ka{\kappa}
\def\la{\lambda}

\def\RE{\Re\mathfrak{e}}
\def\IM{\Im\mathfrak{m}}

\def\di{\mathrm{d}}
\def\D{\mathrm{D}}
\def\T{\mathrm{T}}

\def\A{\mathcal{A}}

\def\pd{\partial}
\def\ld{{\mathscr{L}}}

\def\la{\langle}\def\ra{\rangle}

\def\tr{\mathrm{\,tr\,}}

\def\to{\rightarrow}

\def\ii{\mathrm{i}}

\def\dif{\partial}

\def\ga{\gamma}
\def\de{\delta}
\def\De{\Delta}
\def\ep{\epsilon}
\def\lam{\lambda}
\def\la{\lambda}
\def\rh{\rho}
\def\si{\sigma}
\def\ka{\kappa}

\def\ECM{E_{\text{cm}}}
\def\EB{\bar{E}_{\text{cm}}}

\def\cut{\Lambda_{\textrm{UV}}^{}}

\def\T{{\mathcal{T}}}
\def\C{{\mathrm{C}}}
\def\el{{\mathrm{el}}}
\def\inel{{\mathrm{inel}}}
\def\RE{\Re\mathfrak{e}}
\def\IM{\Im\mathfrak{m}}
\def\CC{\mathcal{C}}

\def\End{\end{document}}

\numberwithin{equation}{section}

\begin{document}

 \thispagestyle{empty}
 \setcounter{footnote}{0}
 \titlelabel{\thetitle.\quad \hspace{-0.8em}}
\titlecontents{section}
              [1.5em]
              {\vspace{4mm} \large \bf}
              {\contentslabel{1em}}
              {\hspace*{-1em}}
              {\titlerule*[.5pc]{.}\contentspage}
\titlecontents{subsection}
              [3.5em]
              {\vspace{2mm}}
              {\contentslabel{1.8em}}
              {\hspace*{.3em}}
              {\titlerule*[.5pc]{.}\contentspage}
\titlecontents{subsubsection}
              [5.5em]
              {\vspace{2mm}}
              {\contentslabel{2.5em}}
              {\hspace*{.3em}}
              {\titlerule*[.5pc]{.}\contentspage}

\begin{center}
{\bf {\Large
Unitary Standard Model from Spontaneous Dimensional Reduction
\\[1.5mm]
and Weak Boson Scattering at the LHC}}

\vspace*{8mm}

{\sc Hong-Jian He}\,$^{a,b,c}$ ~~and~~ {\sc Zhong-Zhi Xianyu}\,$^a$

\vspace*{3mm}

$^a$\,Institute of Modern Physics and Center for High Energy Physics,
\\
Tsinghua University, Beijing 100084, China
\\[1mm]
$^b$\,Center for High Energy Physics, Peking University, Beijing 100871, China
\\[1mm]
$^c$\,Kavli Institute for Theoretical Physics China, CAS, Beijing 100190, China
\\

\vspace*{25mm}
\end{center}

\vspace*{3mm}

\begin{abstract}
\baselineskip 17pt
\noindent
Spontaneous dimensional reduction (SDR) is a striking phenomenon predicted
by a number of quantum gravity approaches which all indicate that
the spacetime dimensions get reduced at high energies. In this work,
we formulate an effective theory of electroweak interactions based upon
the standard model, incorporating the spontaneous reduction of
space-dimensions at TeV scale. The electroweak gauge symmetry is
nonlinearly realized with or without a Higgs boson.
We demonstrate that the SDR ensures good high energy behavior and
predicts unitary weak boson scattering.
For a light Higgs boson of mass 125\,GeV, the TeV-scale SDR
gives a natural solution to the hierarchy problem.
Such a light Higgs boson can have induced anomalous gauge couplings from
the TeV-scale SDR.  We find that the corresponding $WW$ scattering cross sections
become unitary at TeV scale, but exhibit different behaviors from that of the
4d standard model. These can be discriminated by the $WW$ scattering experiments
at the LHC.
\\[4mm]
%
%
\hfill  Eur.\ Phys.\ J., in Press [\,arXiv:1112.1028\,]
\end{abstract}

\newpage
\setcounter{page}{2}

\tableofcontents

 \setcounter{footnote}{0}
 \renewcommand{\thefootnote}{\arabic{footnote}}

\baselineskip 18.5pt

\vspace*{10mm}
\section{Introduction}
\vspace*{1.5mm}

 So far it is well established both theoretically and experimentally
 that the electroweak interactions fit in the gauge structure of
 $\,SU(2)_L^{}\otimes U(1)_Y^{}\,$,\, and are mediated by the massless photon
 plus three massive weak gauge bosons $(W^\pm,Z^0)$.\,
 In the electroweak standard model (SM) \cite{SM},
 the gauge symmetry $\,SU(2)_L^{}\otimes U(1)_Y^{}\,$ is linearly realized and
 spontaneously broken by the conventional Higgs mechanism \cite{HM}
 in 4-dimensions. A Higgs boson is predicted and is crucial for the renormalizability
 \cite{tHooft-Veltman} and unitarity \cite{SMunitary,SMunitary1}
 of the SM.  Recently, the ATLAS and CMS experiments at the LHC
 have found a new particle with mass around 125\,GeV,
 whose signals differ from that of the SM Higgs boson in the diphoton channel,
 but are still consistent with the SM (within about $2\sigma$ statistics)
 \cite{LHCnew,Moriond}.
 Therefore, further LHC explorations are important to pin down
 the true mechanism of the electroweak symmetry breaking, and
 the possible new physics beyond a na\"ive SM Higgs boson is highly anticipated.
 Since so far the LHC has found no other new particles,
 it is pressing to explore alternative new physics sources
 beyond the traditional proposals (such as extra dimensions, supersymmetry
 and strong dynamics) at the weak scale.  Different ways
 of unitarizing the longitudinal weak boson scattering around TeV scale reflect
 the corresponding different underlying mechanisms of electroweak symmetry breaking (EWSB),
 and will be discriminated by the $WW$ scattering experiments as a key task
 of the CERN LHC \cite{WW-rev}.

As is known, na\"ively introducing a bare mass-term of weak gauge bosons or
SM fermions into the Lagrangian \emph{by hand} would ruin both the renormalizability
and unitarity of the theory.
Such bare mass-terms can be made gauge-invariant via nonlinearly realized
electroweak gauge symmetry in the minimal Higgsless SM, but suffer strong
unitarity limits for the $W/Z$ and the SM fermions, all within the energy range of
$\,1.2-170\,$TeV \cite{SMunitary,SMunitary1,Chanowitz,Dicus:2004rg}.
The SM resolves this problem by invoking a Higgs doublet
to linearly realize the electroweak gauge symmetry and its spontaneous
breaking via Higgs vacuum expectation value.   With this, the weak gauge bosons $(W,Z)$
as well as SM fermions acquire the observed masses, and a physical Higgs boson is predicted.
Such a theory is indeed renormalizable\,\cite{tHooft-Veltman} and
unitary\,\cite{SMunitary,SMunitary1,Chanowitz,Dicus:2004rg}
as ensured by the presence of the physical Higgs boson, and is usually expected to be
ultraviolet (UV) complete up to high energies below the Planck scale
$\MP\,(\simeq 1.2\times 10^{19}$\,GeV).\,
However, such a fundamental SM Higgs boson suffers a number of theoretical
inconsistencies, including the unnaturalness\,\cite{unnatural} and
triviality\,\cite{trivial}.
To overcome these difficulties, many candidates have been put on market,
incorporating new physics such as compositeness\,\cite{strong},
supersymmetry\,\cite{susy}, and extra dimensions\,\cite{extrad}
or deconstructions\,\cite{DC}.

In this work, we present a novel, but conceptually simple and attractive
approach to the electroweak symmetry breaking via spontaneous dimensional reduction
(SDR). In our construction, the electroweak gauge symmetry
is nonlinearly realized with or without a Higgs boson,
while the renormalizability and unitarity will be retained,
due to spontaneous reduction of space-dimensions in the high energy regime.
The concept of dimensional reduction was conceived
by 't\,Hooft\,\cite{thooft} in the context of holographic principle,
and the phenomenon of SDR has been predicted by a number of quantum gravity approaches.
Some well-known and intriguing studies of quantum gravity along this direction
include\,\cite{carlip} such as the exact renormalization group approach\,\cite{ERG},
the causal dynamical triangulation \cite{ambjorn},
the loop quantum gravity\,\cite{LQG}, the high-temperature string theory\,\cite{HTstring},
and Ho\v{r}ava-Lifshitz gravity \cite{horava}.
(The possibility of dynamical generation of the space-dimension(s) at low energies
has also been stressed and explored via dimensional (de)construction\,\cite{DC}.)
Despite the differences in their detailed structures,
all these studies\,\cite{carlip} reveal a {\it common feature,}
showing that the spacetime dimensions, being $\,n=4\,$ in the infrared region,
will become reduced and continuously approach $\,n=2\,$ in the UV limit,
due to nonperturbative quantum gravity corrections.
This essential feature is truly appealing,
since in general a field theory exhibits worse UV behavior in
higher spacetime dimensions than 4d, but displays better UV behavior for lower dimensions.
For instance, a gauge theory (including the SM) in $\,n\geqq 5\,$ dimensions
is superficially nonrenormalizable. On the contrary,
the SDR predicts the spacetime dimension to decrease towards $\,n=2\,$ in the UV region,
and will thus substantially improve the UV behavior of the theory.
At the present, the high energy behavior of the SM is far from being well tested and understood.
With the LHC running in its first phases of \,$7-8$\,TeV,\, we are just entering the TeV scale.
The decisive measurements of the longitudinal weak boson scattering\footnote{%
For convenience, we simply use the symbol $W$ to denote both the weak bosons
$W^\pm$ and $Z^0$ in this paper, unless specified otherwise.}
$\,W_LW_L\to W_LW_L\,$
and thus the probe of the UV behavior as well as its underlying EWSB mechanism
have to rely on the second phase of the LHC runs at \,$13-14$\,TeV.\,
This also means that the energy scale at which the SDR becomes significant
can be as low as TeV scale \cite{mureika}.
--- In this work, we conjecture that the TeV Scale SDR can play a key role to
make the Higgsless or Higgsful SM unitary and renormalizable.
This will be definitively tested by the $WW$ scattering experiments at the LHC.

In spite of no complete theory of quantum gravity available that could precisely
determine the SDR, we will approach this problem via the
{\it effective theory formulation} \cite{EFT}.
Thus, we will parameterize the spacetime dimension $\,n=n(\mu)\,$ as a smooth function
of the energy scale $\,\mu\,$ (called the dimensional flow \`{a} la
Calcagni\,\cite{calcagni}).
The $\,n(\mu)\,$ behaves as $\,n(\mu)\to 4\,$ under $\,\mu\to 0\,$
in the infrared region for the agreement with all low energy experiments,
and $\,n(\mu )\to 2\,$ at an UV scale $\,\cut\,$,
which serves as the UV cutoff of our effective theory.
This will mimic the numerical predictions of
causal dynamical triangularization \cite{ambjorn}.
For instance, we can choose a simple form of the dimensional flow,
\bge
\label{eq:DFansatz}
n(\mu) ~=~ 4-2\left(\!\FR{\mu}{\,\cut\,}\!\right)^{\! \gamma}\,,
\hspace*{15mm} (\,\mu\leqq \cut\,)\,,
\ede
where the index $\,\gamma > 1\,$ is a model-dependent parameter, arising from the
nonperturbative dynamics of quantum gravity.
Other variations\,\cite{ambjorn,calcagni} of (\ref{eq:DFansatz})
are possible before finding a unique full theory of quantum gravity,
but this does not change the main physics features of our study.
In this work, we will show that, if the SDR indeed occurs at the TeV scale,
then an effective theory can be consistently constructed
for scales below $\,\cut$\,.\,
Here the new physics effect will enter at low energy scales $\,\mu < \cut\,$
through the modification of $\,n(\mu)\,$ in (\ref{eq:DFansatz}).
Practically, we will take $\,\cut =\order{5\,\text{TeV}}$,\,
as required to ensure the unitarity of $WW$ scattering.
Our first construction is the minimal Higgsless SM with SDR (HLSM-SDR).
It contains no SM Higgs boson,
yet it remains manifestly renormalizable and unitary at scales below $\,\cut$\,.\,
Even though our model differs from all other conventional candidates for the EWSB and
the new physics at the TeV scale, it does coincide with the intuitions and predictions
of many quantum gravity theories\,\cite{carlip,ERG,ambjorn,LQG,HTstring,horava},
showing that the UV behavior of a field theory
will be improved by lowering the spacetime dimensions.
We also note that a SM model without the Higgs boson is still consistent
with the current experimental results, where the recently observed
125\,GeV boson\,\cite{LHCnew} can be something else, such as a dilaton-like
particle \cite{dilaton,Antipin:2013kia}.

For the second construction, we consider the Higgsful SM under the SDR (HFSM-SDR).
We note that the quantum-gravity-induced SDR at TeV scales can
provide a natural solution to the gauge hierarchy problem
(naturalness problem) \cite{unnatural} that plagues
the conventional 4d SM with a Higgs boson.
For the TeV-scale SDR, new physics effects arising from quantum gravity
are expected in the low energy effective theory.  This means that
the Higgs boson can be non-SM-like and encode such new physics in
its anomalous gauge couplings with $WW$ and $ZZ$, as well as in its self-couplings.
For the conventional 4d SM with a non-standard Higgs boson\,\cite{HVV-0},
it was found\,\cite{He:2002qi} that the $WW$ scattering has non-canceled $E^2$ behavior
in the TeV range, and can be probed at the LHC \cite{He:2002qi,HVV-2}.
But additional new physics is required
to unitarize the non-canceled $E^2$ contributions to the $WW$ scattering.
Our current work studies the new construction of HFSM-SDR,
where the light Higgs boson of mass 125\,GeV has anomalous couplings
induced from the TeV scale SDR.
We demonstrate that under the SDR the corresponding $WW$ scattering cross sections become
unitary at TeV scale, but exhibit different behaviors at the LHC.

 This paper is organized as follows.
 In Sec.\,2, we present the construction of the SM incorporating the SDR.
 Then,  we consider the HLSM-SDR in Sec.\,3 as a simple example,
 in which the effect of SDR is easily seen.
 We apply the model to study the $WW$ scattering, and demonstrate
 how the unitarity is guaranteed by the SDR.
 In Sec.\,4 we study the HFSM-SDR containing a non-standard Higgs boson
 with anomalous gauge couplings, and show that the non-canceled $E^2$
 contributions to the $WW$ scattering get unitarized by the SDR at the TeV scale.
 We demonstrate that both HLSM-SDR and HFSM-SDR can be discriminated
 from the conventional unitarization schemes via $WW$ scattering at the LHC.
 We finally conclude in Sec.\,5.
 In Appendix-A, we compute the final-state phase space in general $n$-dimensions,
 and derive the corresponding unitarity bounds for both partial wave amplitudes
 and cross sections.

\vspace*{3mm}
\section{Spontaneous Dimensional Reduction for the SM}
\vspace*{1.5mm}

 We present an effective theory description of SDR,
 which encodes the information of dimensional flow
 $\,n=n(\mu)\,$ into the measure of the spacetime integral $\,\di\rho\,$,\,
 and use $\,\di\rho\,$ to replace all integral measure $\,\di^4x\,$ appearing
 in the action functional.  Ref.\,\cite{calcagni} proposed a
 rigorous mathematical construction of $\,\di\rh\,$,\,
 but it is not needed for the present study at tree level.
 All what we need to know is that this measure has the mass-dimension given by
 $\,[\di\rh]=-n\,$,\, with $\,n=n(\mu)\,$ the dimensional flow introduced in Sec.\,1.
 For this purpose, it suffices to define the measure $\,\di\rh\,$ formally
 by $\,\di^nx\,$,\, and the quantity $\,n\,$ is scale-dependent.
 Then, the action of the theory can be written as,
 \beqa
 \label{eq:L}
   S ~= \int\!\di^nx\,\ld
   ~= \int\!\di^nx\, \(\ld_G^{} + \ld_F^{}\)   \,,
 \eeqa
 where $\,\ld_G^{}\,$ and $\,\ld_F^{}\,$ represent
 the gauge and fermion parts of the SM Lagrangian.
 It has been shown by Calcagni \cite{calcagni} that the action can also be constructed
 in such a ways that the Lagrangian density $\,\ld$\, lies in $3+1$ dimensional spacetime
 and respects the $(3+1)$d Poincar\'e symmetry, while the effect of SDR is fully governed
 by a properly defined integral measure $\di\rho$\,.\,
 In such a scenario, scalar, spinor and vector fields are linear representations of
 (3+1)d Lorentz group $SO(3,1)$ (up to a gauge transformation for gauge fields).
 Practically, this is similar to the conventional dimensional reduction regularization
 method \cite{DRED}, which maintains the 4d Lorentz symmetry and continues physics to
 $\,n < 4\,$.\,
 Alternatively, one can also define the theory in $\,n\leqq 4$\, dimensions
 under the prescription of conventional dimensional regularization method \cite{DREG}
 which continues the Lagrangian to $\,n<4\,$
 and the Lorentz group to $\,SO(n\!-\!1,1)\,$.\,
 We note that either of the two prescriptions leads to the same result for our study
 of longitudinal $WW$ scattering. This is because such longitudinal scattering amplitudes
 equal that of the corresponding Goldstone bosons at high energies
 based upon the equivalence theorem \cite{ET}.
 The manipulations of scalar fields do not invoke
 contracting or counting Lorentz indices, so details of the SDR realizations are
 irrelevant here.
 This will be explicitly justified by the calculations of the next section.

 In the present work, we will focus on the gauge sector,
 \beqa
 \label{eq:L-G}
   \ld_G^{} ~=\, -\fr{1}{4}W_{\mu\nu}^aW^{\mu\nu a}_{} - \fr{1}{4}B_{\mu\nu}^{}B^{\mu\nu}_{}
           +M_W^2 W_\mu^+W^{-\mu}+ \fr{1}{2}M_Z^2Z_\mu^{} Z^\mu_{} \,,
 \eeqa
 with
 \beqs
 \label{eq:W-F}
 \beqa
 \label{eq:Wmunu}
   W_{\mu\nu}^a &\,=\,& \pd_\mu^{} W_\nu^a-\pd_\nu^{} W_\mu^a + g\ep^{abc}W_\mu^b W_\nu^c \,,
   \\[1mm]
 \label{eq:Fmunu}
   B_{\mu\nu}^{} &\,=\,&  \pd_\mu^{} B_\nu^{}-\pd_\nu^{} B_\mu^{} \,,
   \\[1mm]
   M_Z^{} &\,=\,& \frac{M_W^{}}{\,\cos\theta_w^{}\,} \,,
 \eeqa
 \eeqs
 where $(a,b,c)$ denote the $SU(2)$ gauge-indices, and
 $\,\theta_w=\arctan (g'/g)\,$ is the weak mixing angle connecting gauge-eigenbasis
 \,$(W^3_\mu,\,B_\mu^{})$\, to the mass-eigenbasis \,$(Z^0_\mu,\,A_\mu^{})$\,.\,
 The mass-eigenbasis fields
 $\,W_\mu^\pm\,$,\, $Z_\mu^0$\, and $\,A_\mu$\, fields are defined as usual,
 \beqs
 \beqa
   W_\mu^\pm &\,=\,& \fr{1}{\sqrt2}(W_\mu^1\pm\ii W_\mu^2) \,,
   \\[2mm]
   {Z_\mu^0\choose A_\mu} &\,=\,&
   \begin{pmatrix}
   \cos\theta_w & -\sin\theta_w
   \\[1.5mm]
   \sin\theta_w & \cos\theta_w
   \end{pmatrix}\! {W_\mu^3\choose B_\mu} \,.
 \eeqa
 \eeqs
 For the fermion sector, let us consider a pair of generic SM fermions
 \,$(\psi_1^{},\,\psi_2^{})$.\, They form a left-handed $SU(2)_L^{}$ doublet
 $\Psi_L^{}=(\psi_{1L}^{},\,\psi_{2L}^{})^T$\, (with hypercharge $Y_L^{}$), and
 two right-handed weak singlets $\,\psi_{1R}^{}\,$ and $\,\psi_{2R}^{}\,$
 (with hypercharges $Y_{1R}^{}$ and $Y_{2R}^{}$).
 Hence, \,$(\psi_1^{},\,\psi_2^{})$\, have electric charges
 $\,Q_1^{}=\fr{1}{2}+Y_L^{}=Y_{1R}^{}\,$ and
 $\,Q_2^{}=-\fr{1}{2}+Y_L^{}=Y_{2R}^{}\,$,\, respectively. Then, we can write down
 the bare Dirac mass-terms for \,$(\psi_1^{},\,\psi_2^{})$\,,
 \beqa
 \label{eq:L-Fmass}
 \ld_{Fm}^{} ~=\, - m_1^{}\bar{\psi}_1^{}\psi_1^{}
                 - m_2^{}\bar{\psi}_2^{}\psi_2^{} \,.
 \eeqa

 The Lagrangians (\ref{eq:L-G}) and (\ref{eq:L-Fmass})
 derive directly from the SM in the unitary gauge and with Higgs boson removed.
 We note that (\ref{eq:L-G}) is just the lowest order of the
 SM electroweak chiral Lagrangian with nonlinearly realized
 $\,SU(2)_L^{}\otimes U(1)_Y^{}\,$ symmetry in the unitary gauge \cite{App}.
 To make this clear, we can rearrange the Higgs doublet $\,\phi$\,,\,
 together with its charge conjugation field $\,\tilde\phi=\ii\tau^2\phi^*\,$
 into a $2\times 2$ matrix
 $\,\Phi=(\tilde\phi,\,\phi)$\,,\, which can be parameterized as follows,
 \beqa
 \label{eq:Phi-Sigma}
   \Phi ~=~ \fr{1}{2}(v+h)\Sigma \,, &~~~~~&
   \Sigma ~=~ \exp [\ii\tau^a\pi^a/v] \,,
 \eeqa
 where $\tau^a~(a=1,2,3)$ are Pauli matrices. Then, to the lowest order of derivative expansion\,\cite{wein79}, we have the following
 kinematic and interaction terms for $\Sigma$ field and
 the singlet Higgs field $h$\,,
 \beqa
 \label{eq:L-Sigma}
 \ld_H^{} &\!=\!&
 \frac{1}{4}\(v^2+2\ka v h +\ka' h^2\)
 \tr\!\!\left[(\D^\mu\Sigma)^\dag(\D_\mu\Sigma)\right]
 \n\\[1mm]
 &&
 +\frac{1}{2}\dif_\mu^{}h\dif^\mu h
 -\frac{1}{2}M_h^2h^2 - \frac{\lambda_3^{}}{3!}vh^3
 +\frac{\lambda_4^{}}{4!}h^4 \,,~~~~~
 \eeqa
 as formulated before \cite{HVV-0}\cite{He:2002qi} for the 4d non-SM Higgs boson.
 The covariant derivative in (\ref{eq:L-Sigma}) is defined as
 \beqa
 \label{eq:D-Sigma}
   \D_\mu\Sigma ~=~
   \pd_\mu^{}\Sigma
   + \fr{\ii}{2} g{W}_\mu^a\tau^a\Sigma-\fr{\ii}{2}g'B_\mu\Sigma\,\tau^3 \,.
 \eeqa
 It is clear that for the unitary gauge $\,\Sigma=1\,$,\,
 the above kinematic term directly reduces to
 the mass terms of $W^\pm$ and $Z^0$ bosons
 with $\,M_W^{} = \frac{1}{2}gv\,$,\, as shown in (\ref{eq:L-G}).
 Under the SM gauge group $\,SU(2)_L^{}\otimes U(1)_Y^{}\,$,
 the matrix $\Sigma$ transforms as,
 \beqa
 \label{eq:GT}
 \Sigma &\to& \Sigma' ~=~ U_L^{}\Sigma \,U_Y^\dag \,,
 \eeqa
 and the singlet $\,h\to h'=h\,$,\,
 where $\,U_L^{}=\exp[-i\theta_L^a\tau^a/2]\in SU(2)_L^{}\,$ and
 $\,U_Y^{}=\exp[-i\theta_Y^a\tau^3/2]\in U(1)_Y^{}\,$.\,
 We see that the nonlinear Lagrangian (\ref{eq:L-Sigma}) is gauge-invariant under the
 transformation (\ref{eq:GT}). Since (\ref{eq:L-Sigma}) induces the bilinear gauge-Goldstone
 mixing terms for $\,W_\mu^\pm-\pi^\mp\,$ and $\,Z_\mu^0-\pi^0\,$,\, we can impose the
 $R_\xi^{}$ gauge-fixing terms as usual,
 \beqa
 \label{eq:Rxi}
 && \hspace*{-8mm}
 \ld_{\text{gf}}^{} ~=\, -\frac{1}{\xi_W^{}}F_+F_-
                            -\frac{1}{2\xi_Z^{}}F_Z^2
                            -\frac{1}{2\xi_A^{}}F_A^2 \,,
 \\[1.5mm]
 && \hspace*{-8mm}
 F_\pm^{} \,=\, \partial^\mu W^\pm_\mu -\xi_W^{}M_W^{}\pi^\pm , ~~~~~
 F_Z^{} \,=\, \partial^\mu Z^{}_\mu -\xi_Z^{}M_Z^{}\pi^0 , ~~~~~
 F_A^{} \,=\, \partial^\mu A^{}_\mu \,,
 \n
 \eeqa
 and the corresponding Faddeev-Popov ghost terms.
 So, in this gauge all gauges bosons have well-behaved propagators which scale with
 the momentum as $\,1/p^{2}\,$ in the high energy limit.

 For completeness, we can also use $\,\Sigma\,$ field to
 rewrite the fermion mass-term (\ref{eq:L-Fmass})
 into the gauge-invariant form,
 together with the leading order Yukawa interactions,
 \beqa
 \label{eq:L-FmassSigma}
 \ld_{FH}^{} ~=~  -
 \overline{\Psi_L^{}}\,\Sigma\,
 \left[M_f^{} + \fr{1}{2} {\cal Y}_f^{} h \right]
 \,\Psi_R^{} + \text{h.c.}\,,
 \eeqa
 where $\,\Psi =(\psi_1^{},\,\psi_2^{})^T\,$ and
 $\,M_f^{}=\text{diag}(m_1^{},\,m_2^{})\,$.\, The Yukawa coupling
 $\,{\cal Y}_f^{} = 2M_f^{}/v +\Delta{\cal Y}_f^{}\,$,\, where
 $\,\Delta{\cal Y}_f^{}\,$ denotes the possible anomalous coupling
 beyond the SM.  Further incorporation of Majorana mass terms
 for light neutrinos is given in \cite{Dicus:2004rg}.

 If the Higgs singlet field $h$ is removed from the Lagrangian (\ref{eq:L-Sigma}),
 the remaining term is usually viewed as the leading order Lagrangian
 of a nonlinear effective theory of spontaneous electroweak symmetry breaking
 with strong dynamics, and the gauge-invariant higher order
 nonrenormalizable operators can be systematically formulated\,\cite{App}
 under derivative expansion\,\cite{wein79}.
 These higher order operators will induce unknown
 coefficients order by order and the unitarity cannot be restored unless a UV completion
 model is constructed to provide the proper unitarization.
 This is because the Lagrangians (\ref{eq:L-Sigma}) and (\ref{eq:L-FmassSigma})
 (with Higgs removed or with a non-standard Higgs boson)
 in 4-dimensional spacetime will cause bad UV behavior,
 and thus violate both renormalizability and unitarity.
 However, as will be shown below, once the SDR is implemented, both troubles disappear.
 So we turn to the SDR and explain what new features it will bring to our construction.

 We first recall that the action functional in $n$-dimensional spacetime with any \,$n$\,,\,
 should remain dimensionless. This implies that the Lagrangian has the mass-dimension
 $\,[\ld]=n$\,.\,   In consequence, the gauge coupling $\,g\,$,\,
 which has mass dimension $\,[g]=(4-n)/2\,$,\, becomes dimensionful when SDR takes place.
 Thus, we can always define a new coupling $\,\tilde g\,$ which keeps dimensionless at all
 relevant scales,  by transferring the mass-dimension of $\,g\,$ to a proper mass-parameter.
 Since the only dimensionful parameter appearing in the Lagrangian (\ref{eq:L-G})
 at dimension-4 is the $W$ mass $M_W$, it is reasonable to define the dimensionless
 $\,\tilde g\,$ as follows,
 \beqa
 \label{eq:gt}
   g ~=~ \tilde g\, M_W^{(4-n)/2} \,,
 \eeqa
 and the value of the dimensionless coupling $\,\tilde g\,$ is given by that of $\,g\,$
 at dimension $\,n=4\,$.\, This prescription is a part of the definition
 of our model, and we will concentrate on the tree-level analysis for the present study.

 In fact, the scaling (\ref{eq:gt}) is well justified for more reasons.
 We may easily wonder why we could not use the UV cutoff $\cut$ in the scaling of $g$ as
 a replacement of the infrared mass-parameter $M_W$ of the theory.
 This is because that in spacetime dimension $\,n<4$\,,\,
 the gauge coupling $g$ is super-renormalizable with positive mass-dimension
 $\,[g]=(4-n)/2 > 0\,$.\,
 Such a super-renormalizable coupling must be insensitive
 to the UV cutoff of the theory, contrary to a non-renormalizable coupling
 with negative mass-dimension
 ({\it e.g.,} in $n>4$,\, or in association with certain higher-dimensional operators)
 and naturally suppressed by negative powers of the UV cutoff $\cut$.
 It is easy to imagine that for a super-renormalizable theory
 in dimension $\,n<4\,$,\, if its coupling $\,g\,$ were scaled as
 $\,g=\tilde{g}\Lambda_{\rm UV}^{(4-n)/2}$,\,
 it would even make tree-level amplitude UV divergent and blow up as $\cut\to\infty$;
 this is clearly not true. On the other hand, it is well-known that a non-renormalizable
 coupling $g$ with negative mass-dimension $\,[g]\equiv -p <0$\,
 should be scaled as $\,g=\tilde{g}/\Lambda_{\rm UV}^{p}\,$,\,
 and thus the tree-level amplitude
 naturally approaches zero when $\,\cut\to\infty$,\, as expected.\footnote{For
 nonrenormalizable theories such as a massless 5d Yang-Mills theory without
 compactification, it was explicitly shown\,\cite{SekharChivukula:2001hz}
 that the unitarity violation of the gauge boson scattering occurs at
 a scale of ${\cal O}(1/g_5^{2})$ [cf.\ Eq.\,(6) of Ref.\,\cite{SekharChivukula:2001hz}],
 where the 5d gauge coupling $g_5^{}$
 has a mass-dimension $-\hf$,\, and the UV cutoff scale $\cut$ of this
 nonrenormalizable theory was naturally identified to be at the order of $1/g_5^{2}$,
 i.e. $\,\cut \sim 1/g_5^{2}\,$ \cite{SekharChivukula:2001hz},
 or one can define $\,g_5^2 =\bar{g}^2/\cut\,$ with $\bar{g}$ being dimensionless.
 Then, {\it after compactification,} it was noted \cite{SekharChivukula:2001hz} that
 this theory contains one more scale in the infrared
 which makes a nontrivial distinction, namely,
 the compactification scale $1/R$,\, which characterizes the mass scale $M_1$ of the lightest
 Kaluza-Klein (KK) state. It is thus a well-known fact that the compactified 4d KK
 theory has a dimensionless gauge coupling $\tilde{g}$ related to the 5d coupling via,
 $\,g_5^2=\tilde{g}^2(\pi R)\sim \tilde{g}^2/M_1\,$, where the lightest
 KK gauge boson mass $\,M_1\sim 1/R$\, is an infrared mass-parameter.}

 Now we can easily realize that if spacetime dimension flows to $\,n=2\,$
 in the UV limit, then the theory defined by (\ref{eq:L}) is well-behaved at high energies.
 This is because all gauge couplings of the Lagrangian (\ref{eq:L}) in $\,n<4\,$ dimensions
 becomes super-renormalizable, and the gauge boson propagators scale as $\,1/p^{2}\,$
 in high momentum limit under the $R_\xi^{}$ gauge-fixing (\ref{eq:Rxi}).
 The only concern is the nonlinear Lagrangians
 (\ref{eq:L-Sigma}) and (\ref{eq:L-FmassSigma}) for the gauge boson and fermion mass terms
 (after removing the Higgs boson or with non-standard Higgs couplings).
 In 4d this is the origin of nonrenormalizability and unitarity violation, as the expansion
 of matrix $\,\Sigma\,$ gives rise to infinite number of higher dimensional operators
 (involving the Goldstone fields $\pi^a$) whose couplings have negative mass-dimensions.
 But, in our construction the spacetime dimension flows to $\,n=2\,$ in high
 energy limit where the Goldstone fields \,$\pi^a$\, have zero mass-dimension and
 the same is true for the vacuum expectation value (Goldstone decay constant) $\,v\,$.\,
 In fact for \,$n=2$\, dimensions the Lagrangians (\ref{eq:L-Sigma}) and (\ref{eq:L-FmassSigma})
 just describe a 2d gauged nonlinear sigma model including ferimons.
 As is well-known, in 1+1 dimensions, (\ref{eq:L-Sigma}) is renormalizable and
 (\ref{eq:L-FmassSigma}) is super-renormalizable, where the
 exclusion or inclusion of the Higgs boson does not matter
 since all scalar fields are dimensionless in 2d.


 We also note that there exist new mechanisms
 in $\,n<4\,$ dimensions to generate masses for gauge bosons other than the Higgs mechanism.
 For instance, consider the $1+1$ dimensional QED, known as Schwinger model \cite{schwinger}.
 The radiative corrections to the vacuum polarization from
 a massless-fermion loop can contribute a finite nonzero mass
 to the gauge boson\,\cite{schwinger},
 \beqa
 m_\ga^2 ~=~ \frac{e^2}{\pi}\,,
 \eeqa
 where $e$ is the dimension-1 QED gauge coupling.  This is quite expected, because
 in 2d a massless vector particle has no propagating degrees of freedom. Thus,
 the possible physical degrees of freedom for a vector boson in 2d, if exist, can only
 be longitudinal polarization, and the associated vector boson must be massive.
 For the $2+1$ dimensional gauge theories such as the 3d QED, the Chern-Simons term
 also introduces a topological mass to the corresponding gauge field \cite{deser},
 \beqa
 m_{\text{cs}} ~=~ \ka \,e^2 \,,
 \eeqa
 where $\,\ka\,$ is the dimensionless Chern-Simons coupling
 and $\,e\,$ is again the QED gauge coupling with dimension $\frac{1}{2}$\,.\,
 Hence, it is quite natural to have an explicit mass-term for the vector field
 in a lower dimensional field theory. We also note that in these two examples,
 the dimensionful gauge couplings are always proportional to the masses of gauge bosons.
 This also supports our prescription for (\ref{eq:gt}).
 In the next section, we will further demonstrate
 that such a mass term is indeed harmless for the unitarity of high energy scattering of
 longitudinal gauge bosons, due to the onset of SDR.

 Next, we turn to the issue of the dimensional flow $\,n=n(\mu)$\,.\,
 As commented in Sec.\,1, to derive an explicit form of $\,n=n(\mu)\,$ requires to know
 a complete theory of quantum gravity which is not yet available so far.
 However, various constructions of the quantum gravity have already indicated
 that the dimensional flow $\,n=n(\mu)\,$ effectively approaches $\,n=2\,$
 at a UV scale $\,\mu=\cut$\,.\, An easy choice for $\,\cut\,$ would be the Planck scale
 $\,\MP\,$.\,  But, this is certainly not the only choice,
 and it is a very interesting and intriguing possibility that the nonperturbative
 dynamics of the quantum gravity drives $\,\cut\,$
 down to TeV scale\,\cite{mureika}.
 In this case, a number of difficulties associated with the EWSB and $W/Z$
 mass-generations in the SM can be resolved without introducing
 additional {\it ad hoc} hypothetical dynamics.
 As we have explained, the EWSB and mass-generations for $W/Z$ and all SM fermions
 must be tied to the TeV scale \cite{Dicus:2004rg}. So, if the quantum-gravity-induced SDR
 is going to resolve the EWSB and mass-generations for the SM particles, our construction of
 the HLSM-SDR thus provides a support of the TeV scale SDR.

 Hence, it is reasonable and appealing to design a function $\,n=n(\mu)\,$
 such that it mimics the main features of the dimensional flow mentioned above.
 Thus, we conjecture that $\,n=n(\mu)$\, is a smooth monotonic function of the relevant
 energy scale $\,\mu$,\, satisfying
 \beqa
 \label{eq:n2-n4}
   n(0) ~=~4\,, ~~~~\text{and}~~~~n(\cut )~=~2 \,,
 \eeqa
 where $\cut$ is the UV cutoff of our effective theory.
 There certainly exist a lot of functions which fulfill the requirement
 (\ref{eq:n2-n4}) in the effective theory,
 but they should all share the same qualitative physical implications.
 A simple choice is given by (\ref{eq:DFansatz}).
 We will adopt this ansatz as an explicit realization of the dimensional flow
 for the current analysis.
 We note that the SDR itself is an effect induced by quantum gravity,
 which implies that quantum gravity becomes strong and dominant for $\,\mu\geqq\cut\,$,
 where the classical concept of spacetime no longer holds
 and the conventional quantum field theory would break down.
 Therefore, our choice here is only an {\it effective theory} description of
 the spontaneous reduction of spacetime dimensions
 as well as the physical degrees of freedom, as a consequence of nonperturbative
 quantum gravity.  Note that these quantum gravity effects will be significant
 nearby $\cut$, where the concept of spacetime (together with its dimension) can be
 rather different from the conventional one. Hence, for the present study, we will treat
 the spacetime reduction only as an effective theory description of
 the low energy quantum gravity effects.
 In summary, we rewrite our ansatz for the dimensional flow as follows,
 \bge
   \label{eq:DFansatz2}
  n(\mu) ~=~ 4-2\left(\!\FR{\mu}{\,\cut\,}\!\right)^{\! \gamma}\,,
  \hspace*{15mm} (\,\mu\leqq \cut\,)\,,
 \ede
 where $\,\cut\,$ is at the TeV scale, and
 will serve as the UV cutoff of our effective theory.
 The new physics effect will enter at low energy scales
 $\,\mu < \cut\,$ by modifying the dimensional flow $n(\mu)$ as in (\ref{eq:DFansatz2}).
 For practical applications, we will set $\,\cut =\order{5\,\text{TeV}}$\,,
 as required to ensure the unitarity of weak boson scattering in Sec.\,3.
 In (\ref{eq:DFansatz2}),
 the exponential parameter $\,\gamma >1\,$ arises from the nonperturbative dynamics
 of quantum gravity and is thus model-dependent. As the simplest realization
 we may set, $\,\gamma =2\,$ or $\,\gamma =1.5\,$,\, for instance.
 Possible variations\,\cite{ambjorn,calcagni}
 of (\ref{eq:DFansatz2}) are allowed since no unique full theory
 of quantum gravity exists so far, but they will not affect the main physics
 features of our analysis below.
 As another remark, one possible concern may be the potential Lorentz violation around
 the cutoff scale $\cut$. But we note that the Lagrangian $\ld_G+\ld_F$  above
 is manifestly Lorentz invariant and thus does not change the particles' dispersion relations
 from the conventional 4d forms.
 Hence, our model is free from the Lorentz-violation constraints in cosmic ray observations,
 which are derived from modification of photon's dispersion
 with the assumption of a particular form of explicit Lorentz violation.
 Also, as shown by \cite{calcagni},
 the construction of $\di^nx$ in (\ref{eq:L}) may cause a Lorentz violation due to the
 scale-dependence of $n$, but this does not affect the particles' dispersion relations at
 tree-level and thus is irrelevant to our current unitarity analysis.

 Finally, we note that our effective theory formulation is also partly motivated
 by recent studies of asymptotic safety (AS) scheme of quantum general relativity (QGR)
 \`{a} la Weinberg \cite{AS-Wein}\cite{AS-rev}\cite{ERG}.
 The theory is originally defined in (3+1)d in the AS scheme.
 Then, the nonperturbative quantum effects from
 solving the exact renormalization group equation of QGR show
 that the theory will flow to a nontrivial fixed point in the UV limit
 and the graviton two-point function exhibits an
 effective 2d UV behavior\,\cite{AS-rev}.
 The SDR becomes manifest via the anomalous scalings of fields,
 and physical variables such as the spacetime curvature.
 We note that these anomalous scalings
 share the similarity with our effective theory,
 while the fields still live in (3+1)d
 and preserve the (3+1)d Lorentz symmetry.
 As a simplified low energy effective theory, our construction does not
 invoke any detail of the underlying UV dynamics of the AS scenario.
 It is interesting to further study the quantitative connection
 between the SDR and the UV dynamics of the AS scenario.
 Besides, the Ho\v{r}ava-Lifshitz models \cite{horava} of quantum gravity
 also provides a concrete field-theoretical UV-completion with SDR,
 it has explicit and relatively tractable Lagrangian.
 So, it is expected that the various scaling properties mentioned in this section
 will naturally arise from the formulation of this model.
 These two approaches are worth of further consideration for
 model-dependent studies in the future.

\vspace*{3mm}
\section{Unitary Weak Boson Scattering under SDR}

 In the previous section, we have set up the formalism of the SM
 with SDR without or with a Higgs boson.
 We will denote these two cases as HLSM-SDR and HFSM-SDR, respectively.
 In the HLSM-SDR, we can remove the Higgs boson from the conventional 4d SM
 and nonlinearly realize the $\,SU(2)_L^{}\otimes U(1)_Y^{}$\,
 electroweak gauge symmetry.  We further construct the dimensional
 flow of spacetime as a proper function of the relevant energy scale
 to realize the SDR.  The physical picture of this construction is simple enough,
 yet it offers a novel solution to ensure the renormalizability
 and unitarity without a conventional Higgs boson.
 So we will take it as our first example for illustration.
 Such a scenario is still consistent with the current LHC data since the
 125\,GeV boson can be something else, such as a dilaton-like
 particle\,\cite{dilaton,Antipin:2013kia}.
 In this section, we study the scattering
 of longitudinally polarized weak gauge bosons.
 We demonstrate how the weak boson scattering amplitudes are
 unitarized in the HLSM-SDR construction,
 and how the corresponding cross sections can be discriminated
 from the conventional 4d SM at the LHC.

\vspace*{3mm}
\subsection{Analysis of Weak Boson Scattering Amplitudes under SDR}
\vspace*{1.5mm}

 In conventional 4d quantum field theory, the longitudinal polarization
 $\,\ep_L^\mu\,$ of a massive vector boson
 increases with its energy,
 \beqa
 \ep_L^\mu (k) &\,=\,&
 \frac{1}{M}\(|\vec{k}|,\,k^0\vec{k}/|\vec{k}|\)
 ~=~ \frac{~k^\mu\,}{M} + v^\mu(k)\,,
 \\[1.5mm]
 v^\mu (k) &\,=\,& \mathcal{O}\!\(\!\frac{\,M\,}{E}\!\) \!,
 \nn
 \eeqa
 where $M$ is the vector boson mass.
 As a result, the longitudinal scattering amplitude may have dangerous power-law
 dependence on the scattering energy, and thus may cause unitarity violation.

\begin{figure}[t]
   \begin{center}
     \includegraphics[width=0.9\textwidth]{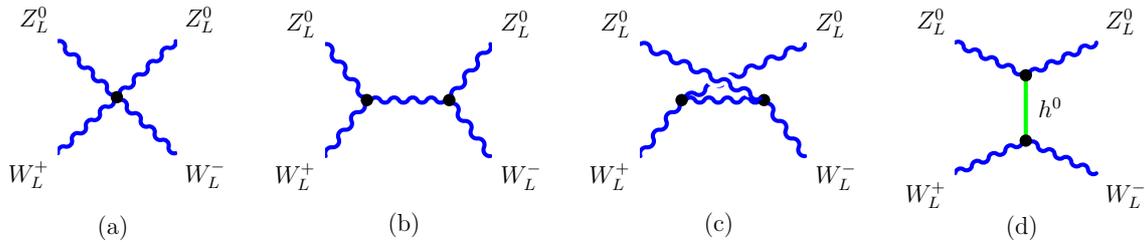}
     \caption{Feynman diagrams for longitudinal scattering $\,W_L^+W_L^-\to Z_L^0Z_L^0$\,
     consist of (a)-(d) in the conventional 4d SM,
     but include only (a)-(c) in the minimal 4d Higgsless SM and in the current HLSM-SDR.}
     \label{Fig-WWZZ}
   \end{center}
 \end{figure}

 The SM Higgs boson plays the key role for ``unitarizing''
 the bad high energy behaviors of
 longitudinal scattering amplitudes\,\cite{SMunitary}.
 To see how this works, let us take an explicit channel of
 $\,W_L^+W_L^-\to Z_L^0Z_L^0\,$.\,
 In the unitary gauge where all would-be Goldstone bosons decouple,
 there are just four diagrams contributing to
 the tree-level scattering amplitude, as shown in Fig.\,\ref{Fig-WWZZ}.
 These diagrams can be readily evaluated in the center-of-mass (c.m.) frame.
 By counting the powers of c.m.\ energy $\,\ECM\,$,\,
 we see that as $\,\ECM\,$ becomes much larger than $W$ boson mass $M_W$,
 the leading energy dependence of the sum of the first three diagrams
 is of $\,\order{\ECM^4}$.\,
 In fact, we can always expand the contribution of each diagram
 by large $\ECM$ expansion, and deduce their sum in the following form,
 \bge
 \mathcal{T} ~=~ A\cdot \ECM^4+B\cdot \ECM^2+\order{\ECM^0} \,.
 \ede
 Unitarity of the $S$-matrix requires vanishing $\,\ECM^4\,$ and $\,\ECM^2\,$ terms, i.e., $\,A=B=0\,$.\,
 The explicit calculation shows that the $\,\ECM^4\,$ terms exactly cancel among the first three diagrams ($\,A=0\,$),
 while summing up the $\,\ECM^2\,$ terms for the first three diagrams yields a nonzero result,
 \bge
   \label{E2term}
  \mathcal{T}_{2}^{}[(a)\!+\!(b)\!+\!(c)] ~=~ \FR{~g^2~}{\,4M_W^2\,}\,\ECM^2\,.
 \ede
  This term is canceled exactly by the $\,\ECM^2\,$ term from the last diagram
  with the SM Higgs boson exchange in the $s$-channel.
  In the conventional 4d SM, it is this cancellation that plays a key role
  to ensure the unitarity for the scattering channel $\,W_L^+W_L^-\to Z_L^0Z_L^0\,$.

 We note that according to the equivalence theorem \cite{ET},
 the longitudinal scattering amplitude equals that of the corresponding
 scalar Goldstone boson scattering at high energies. So such scalar amplitudes
 are cleanly defined for any dimensions.
 Within lower dimensions, the longitudinal amplitude from each diagram remains the same as
 in the 3+1 dimensions.  However, we observe that the condition of the partial wave unitarity
 changes, or more precisely, the form of the partial wave expansion changes,
 {\it due to reduction of phase space in the final state.}
 In consequence, the cancellation of $\,\ECM^2\,$ terms described above
 is no longer necessary anymore.
 This is an essential feature of our Higgsless SM with SDR, i.e.,
 {the $WW$ scattering amplitudes keep unitary at high energies under SDR,
 without invoking a conventional Higgs boson.}

 To be more explicit, let us recall how the unitarity condition is formulated
 in general $n$ dimensions\,\cite{soldate}. In essence, the unitarity condition arises
 from the probability conservation (or information conservation) in a scattering process,
 which imposes the unitarity condition on the $S$-matrix,  $\,SS^\dag=S^\dag S=1$\,,\,
 and thus leads to
 \bge
   \label{UniConT}
   \mathscr{T}^\dag\mathscr{T} ~=~ 2\,\IM \mathscr{T} \,,
 \ede
 where the amplitude $\,\mathscr{T}$\, is defined via $\,S=1+\ii \mathscr{T}$\,.
 We further define the amplitude \,$\T$\,
 via $\,\mathscr{T}=(2\pi)^n\de^n(p_f^{}-p_i^{})\T$\,
 with $\,p_i^{}$\, and \,$p_f^{}$\, the total momenta of initial and final states,
 respectively. For $\,2\to 2\,$ scattering, $\,\T$\, depends only on the c.m.\ energy
 $\,\ECM$\, and the scattering angle $\,\theta\,$.\,
 Thus, in this case  we can always expand $\,\T_\el^{}(\ECM,\theta)$\, in terms
 of partial waves \,$a_\ell^\el(\ECM)$\, for $\,n > 3$\, dimensions,
 %
 \beqa
   \label{PartWaveExp}
   \T_{\text{el}}(\ECM,\theta) ~=~
   \lam_n \ECM^{4-n}\sum_{\ell}\FR{1}{N_\ell^\nu}\C_\ell^\nu(1)
   \C_\ell^\nu(\cos\theta)a^{\text{el}}_\ell(\ECM) \,,
 \eeqa
 where
 \beqa
 \label{eq:la-nu-Nl}
 \lam_n^{} \,=\, 2(16\pi)^{\frac{n}{2}-1}_{}\Gamma(\fr{n}{2}-1)\,,~~~~
 \nu \,=\, \fr{1}{2} (n-3)\,,~~~~
 N_\ell^\nu \,=\, \FR{\pi\Gamma(\ell+2\nu)}{~2^{2\nu-1}\ell!(\ell+\nu)\Gamma^2(\nu)~}\,,
 ~~~~~
 \eeqa
 %
 and $\,\C_\ell^\nu(x)$\, is the Gegenbauer polynomial of order $\,\nu\,$ and degree $\,\ell\,$.\,
 Then, from (\ref{UniConT}) one can derive the unitarity condition for partial waves of
 $\,2\to 2\,$ elastic and inelastic scattering processes,
 \beqa
 \label{eq:UC-PW}
  \big|\Re\mathfrak{e} a_\ell^\el\big| ~\leqq~ \FR{\rh_e^{}}{2} \,, ~~~~~
  \big| a_\ell^\el\big| ~\leqq~ \rh_e^{} \,, ~~~~~
  \big|a_\ell^\inel \big| ~\leqq~ \FR{\,\sqrt{\rh_i^{}\rh_e^{}}\,}{2} ~,
 \eeqa
 where $\,\rh_i^{}\,$ and $\,\rh_e^{}\,$ are symmetry factors associated with the final states,
 for the corresponding $\,2\to 2$\, inelastic and elastic scattering processes, respectively,
 which equal $1!$ ($2!$) for the two final state particles being nonidentical (identical).
 For clarity, we present the derivation of (\ref{eq:UC-PW}) in Appendix\,A.

  We see that the unitarity condition (\ref{eq:UC-PW})
  does not explicitly depend on the spacetime dimension,
  nevertheless, the expression of $\,a_{\ell}^{}\,$ does.
  In fact, for $\,n > 3\,$ dimensions, we can derive,
 \beqa
   \label{eq:PW-nDim}
   a_\ell^\el(\ECM)
   ~=~  \FR{\ECM^{n-4}}{~2(16\pi)^{\fr{n}{2}-1}\Gamma(\fr{n}{2}-1)\C_\ell^{\fr{n-3}{2}}\!(1)~}
       \int_0^\pi\!\di\theta\,\sin^{n-3}\theta \,\C_\ell^{\fr{n-3}{2}}(\cos\theta)\,
       \T_{\text{el}}^{}(\ECM,\theta) \,.~~~~~~~
 \eeqa
 The appearance of the factor $\,\ECM^{n-4}\,$ is expected,
 since the $S$-matrix of $\,2\to 2\,$ scattering
 has a mass-dimension $\,4-n\,$ in $\,n\,$ spacetime dimensions,
 and the partial wave amplitude $\,a_\ell^{}\,$ is dimensionless by definition in general.

 The partial wave expansion in \,$n\leqq 3$\, dimensions needs a separate treatment,
  since for $\,n=3\,$ dimensions, the Gegenbauer polynomial $\,\C_\ell^{(n-3)/2}(x)$\,
  vanishes identically. In fact, the correct eigenfunctions for partial wave expansion in
  $\,n=3\,$ spacetime are simply given by a pure phase $\,e^{\ii\ell\theta}\,$.
 Hence, the correct partial wave expansion in $3$-dimensions reads,
 \beqa
   \label{eq:PW-3D}
   \left\{\begin{split}
     & \T_\el(\ECM,\theta) ~=~ 8\ECM\sum_\ell e^{\ii\ell\theta}a_\ell^\el(\ECM) \,;
   \\
     & a^\el_\ell(\ECM) ~=~
   \FR{1}{\,16\pi\ECM\,}\int_0^{2\pi}\!\di\theta\,e^{-\ii\ell\theta}
   \,\T_{\text{el}}(\ECM,\theta) \,,
   \end{split}\right. ~~~~~~~(\,n=3\,)\,.
 \eeqa
 For the spacetime dimension $\,2\leqq n<3\,$,\,
 it is meaningless to talk about scattering angle and thus no partial wave expansion is needed.
 Or, we would say that the only nonvanishing partial wave here is the $s$-wave $\,a_0^{}\,$,\,
 which can be identified as the forward and backward scattering amplitude multiplied by
 \,$\ECM^{n-4}$\,  to make the partial wave dimensionless,
 \beqa
   \label{eq:PW-2D}
   a^\el_0(\ECM) ~=~
   \FR{\ECM^{n-4}}{\,4^{n-1}\pi^{(n-3)/2}\Gamma(\frac{n-1}{2})\,}
   \left[\,{\T_{\text{el}}(\ECM,0)+\T_{\text{el}}(\ECM,\pi)}\,\right] ,
   ~~~~~~~ (\,2\leqq n<3\,)\,.
 \eeqa
 The coefficients in expansions (\ref{eq:PW-3D}) and (\ref{eq:PW-2D}) are chosen
 such that the $s$-waves obtained from these two equations coincide with that
 from the analytic continuation of (\ref{eq:PW-nDim})
 as a function of the spacetime dimension $n$\,.

 By removing Higgs boson from the SM, we note that the scattering amplitude for
 $\,W_L^+W_L^-\to Z_L^0Z_L^0$\, is dominated by (\ref{E2term}).
 Substituting this into (\ref{eq:PW-nDim}) and (\ref{eq:PW-3D})
 for 4-dimensions and 3-dimensions, respectively,
 and using the unitarity condition (\ref{eq:UC-PW}) for $s$-wave,
 we infer the upper unitary limits on the scattering energy, $\,\ECM < 1.74$\,TeV\,
 in 4-dimensions, and $\,\ECM < 6.02$\,TeV in 3-dimensions.
 But for $\,n=2\,$ dimensions, we should use (\ref{eq:PW-2D}), and
 it yields an energy-independent $s$-wave, \,$a_0^{}=\fr{1}{8}\tilde{g}^2$\,,\,
 which always satisfies the unitarity condition  \,$|a_0^{}|\leqq 1$\,.\,
 So there is simply no unitarity constraint on $\,\ECM$\, in 2-dimensions,
 as we have expected.

 This analysis shows that if SDR is indeed responsible
 for unitarizing the scattering amplitudes of longitudinal weak gauge bosons,
 then the transition scale $\,\cut\,$ should be on the order of TeV scale.
 Otherwise, the unitarity would have been violated before the SDR takes place.
 Actually, we can derive a stronger bound on the transition scale $\,\cut\,$ via
 coupled channel analysis.
 Let us consider the Higgsless Lagrangian (\ref{eq:L-G})-(\ref{eq:W-F}),
 and for current analysis it is enough to set the weak mixing angle
 $\,\theta_w =0\,$ for simplicity.
 For the electronically neutral channels, we have two states
 $\,|W_L^+W_L^-\ra\,$ and  $\,\fr{1}{\sqrt 2}|Z_L^0Z_L^0\ra\,$.\,
 When the Higgs boson is absent, the tree-level amplitude for the process
 $\,Z^0Z^0\to Z^0Z^0\,$ vanishes. Hence, there are only two independent amplitudes
 for evaluation, corresponding to the scattering channels,
 \begin{subequations}
 \label{eq:WW-WW/ZZ}
   \begin{align}
     \label{SingProc1}
     & W_L^+W_L^-\to W_L^+W_L^- \,,\\[1mm]
     \label{SingProc2}
     & W_L^+W_L^-\leftrightarrow Z_L^0Z_L^0 \,.
   \end{align}
 \end{subequations}
 Three diagrams will contribute to the scattering (\ref{SingProc1}),
 via $4W$ contact interaction, $s$- and $t$-channel $Z$ exchanges, respectively.
 We summarize the amplitude for each diagram under the power expansion of the c.m.\
 energy $\,\ECM$\,,
 \begin{subequations}
 \beqa
 \mathcal{T}_{ct}^{} &=&
    \FR{\,g^2}{4}\left[\FR{1}{4}(-3+6\cos\theta+\cos^2\theta )\EB^4
                     + (2-6\cos\theta)\EB^2\right] + \order{\ECM^0} \,,
 \hspace*{14mm}
 \\[1.5mm]
   \mathcal{T}_{Zs}^{}
    &=& \FR{\,g^2}{4}\left[ - \cos\theta\EB^4 - \cos\theta\EB^2\right] +\order{\ECM^0} \,,
 \\[1.5mm]
   \mathcal{T}_{Zt}^{}
    &=& \FR{\,g^2}{4}\left[\frac{1}{4}(3-2\cos\theta-\cos^2\theta)\EB^4
                +\(\!-\frac{3}{2}+\frac{15}{2}\cos\theta\)\EB^2\right] + \order{\ECM^0} \,,
 \eeqa
 \end{subequations}
 where $\,\EB \equiv \ECM /M_W\,$.\,
 We see that the $\,\order{\ECM^4}$\, terms cancel among themselves,
 as guaranteed by the Yang-Mills gauge symmetry.
 But the sum of $\,\order{\ECM^2}\,$ terms gives rise to
 the following nonzero result,
 \beqa
 \label{eq:WWWW-sum}
   \mathcal{T}[W_L^+W_L^-\to W_L^+W_L^-] ~=~
   \FR{g^2\ECM^2}{\,8M_W^2\,}(1+\cos\theta) + \order{\ECM^0} \,.
 \eeqa

  Next, we turn to the process (\ref{SingProc2}).
  The tree-level diagrams for the scattering channel
  $\,W_L^+W_L^-\to Z_L^0Z_L^0\,$ are shown in Fig.\,\ref{Fig-WWZZ}.
  For the absence of Higgs boson,
  only the first three diagrams are relevant,
  and we compute them as follows,
 \begin{subequations}
 \beqa
     \mathcal{T}_{ct} &=&
      \FR{\,g^2}{4}\!\left[\FR{1}{4}(-6+2\cos^2\theta)\EB^4+ 4\EB^2\right] +\order{\ECM^0} \,,
 \\[1.5mm]
     \mathcal{T}_{Wt} &=&
        \FR{\,g^2}{4}\!\left[\FR{1}{4}(3-2\cos\theta-\cos^2\theta)\EB^4
       + \(\!-\FR{3}{2}+\FR{15}{2}\cos\theta\)\EB^2 \right] + \order{\ECM^0} ,
\hspace*{13mm}
 \\[1.5mm]
     \mathcal{T}_{Wu} &=&
      \FR{\,g^2}{4}\!\left[\FR{1}{4}(3+2\cos\theta-\cos^2\theta )\EB^4 +
                         \(\!-\FR{3}{2}-\FR{15}{2}\cos\theta\)\EB^2 \right] + \order{\ECM^0} .
\hspace*{13mm}
 \eeqa
 \end{subequations}
 Again, all the $\,\order{\ECM^4}$ terms sum up to zero,
 and the nontrivial leading amplitude arises at  $\,\order{\ECM^2}\,$,
 \beqa
 \label{eq:WWZZ-sum}
   \mathcal{T} [W_L^+W_L^-\to\fr{1}{\sqrt2}Z_L^0Z_L^0]
   ~=~ \FR{g^2\ECM^2}{\,4\sqrt{2}M_W^2\,} + \order{\ECM^0} \,.
 \eeqa
 With (\ref{eq:WWWW-sum}) and (\ref{eq:WWZZ-sum}),
 we can form the $2\times2$ matrix amplitude
 for the initial/final states $\,|W_L^+W_L^-\ra\,$
 and $\,\fr{1}{\sqrt{2}\,}|Z_L^0Z_L^0\ra\,$,\,
 \begin{align}
   \T_{\text{coup}}^{} ~=~ \FR{\,g^2\ECM^2\,}{\,8M_W^2\,}
   \begin{pmatrix}
   1+\cos\theta & \,\sqrt2\,
   \\[1.5mm]
   \sqrt2 & 0 \end{pmatrix} .
 \end{align}
  \begin{figure}
   \begin{center}
     \includegraphics[width=0.7\textwidth]{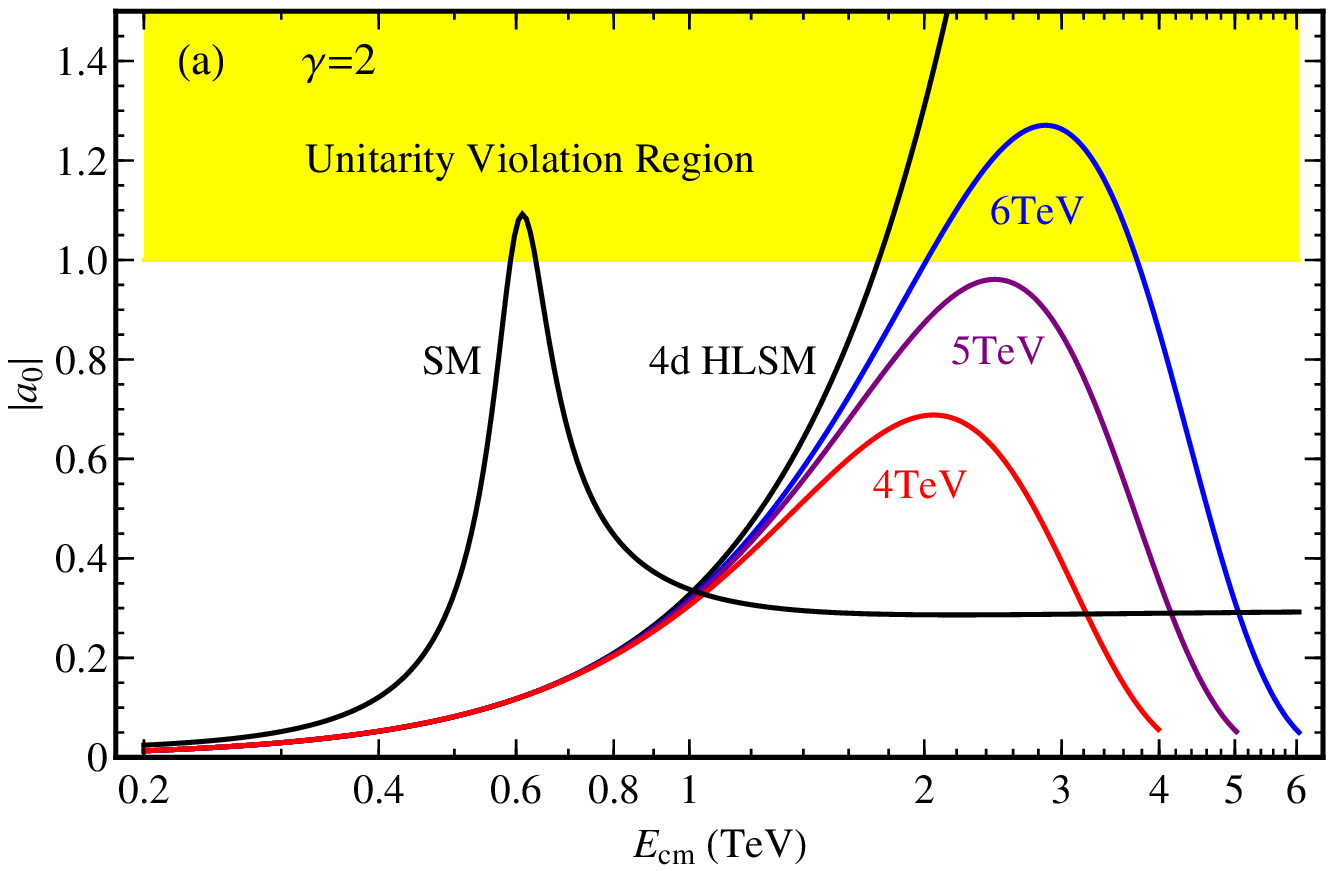}\\[2mm]
     \includegraphics[width=0.7\textwidth]{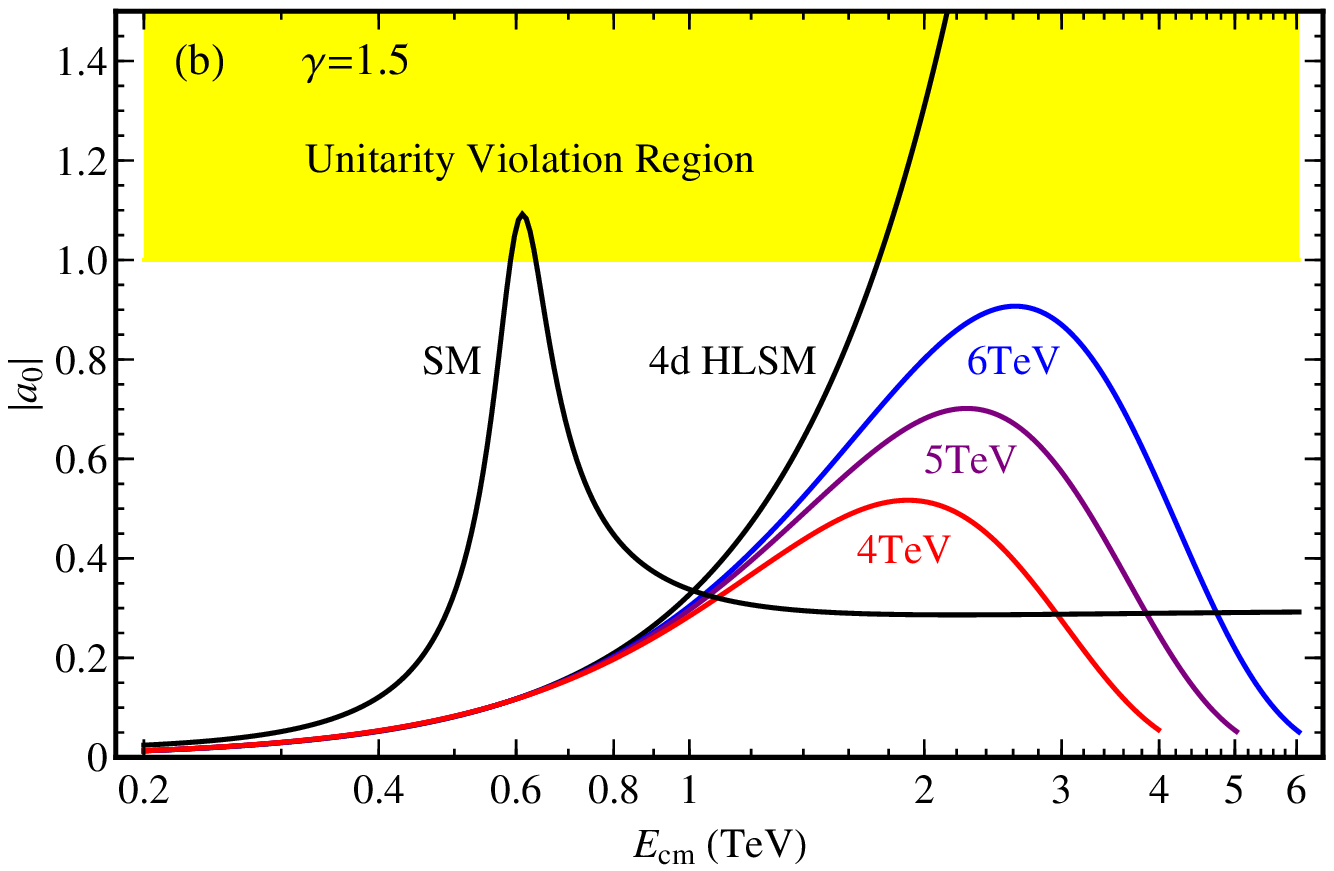}
   \end{center}
   \vspace*{-5mm}
   \caption{Partial wave amplitude of coupled channel scattering as a function of the c.m.\
   energy $\,\ECM$.\,  The $s$-wave amplitudes for the HLSM-SDR
   are shown by red, purple, and blue curves (from bottom to top),
   corresponding to the UV cutoff $\,\Lambda_{\text{UV}}=4,\,5,\,6$\,TeV,
   respectively. Plot-(a) is for index parameter $\,\gamma =2\,$,\, and plot-(b) for
   $\,\gamma =1.5\,$.\,
   As comparisons, the $s$-wave amplitudes for the minimal 4d Higgsless SM (4d\,HLSM)
   and for the conventional 4d SM (with a {600}\,GeV Higgs boson) are also
   shown by the black curves.
   }
   \label{Fig-PartWave}
 \end{figure}

 Then, we derive the $s$-wave amplitude in $n$-dimensions
 with the aid of (\ref{eq:PW-nDim}),
 \begin{align}
 \label{eq:a0-coup}
   a_{0}^{\text{coup}} ~=~
   \FR{\tilde g^2}{\,2^{n+2}_{}\pi^{\fr{n-3}{2}}
   \Gamma(\fr{n-1}{2})\,}\(\FR{\ECM}{2M_W^{}}\)^{\!n-2}\!
   \begin{pmatrix}
   1 & \,\sqrt{2}\,
   \\[1.5mm]
   \sqrt{2} & 0
   \end{pmatrix},
 \end{align}
 where $\,\tilde{g}\,$ is defined by (\ref{eq:gt}).
 We note that the partial wave expansion (\ref{eq:PW-nDim})
 works for $\,n>3$\,,\, and we can analytically continue
 (\ref{eq:a0-coup}) as a function of $\,n$\, to the entire range
 $\,2\leqq n \leqq 4\,$.\, Such an analytic continuation is made in
 the complex plane of spacetime dimension $\,n\,$,\,
 and the one-to-one mapping between $\,n\,$ and $\,\mu\,$ is
 prescribed only within the real interval $\,2\leqq n\leqq 4$\,.\,
 Then, we can diagonalize the matrix in (\ref{eq:a0-coup}),
 \begin{align}
   a_{0}^{\text{diag}} ~=~
   \FR{\tilde g^2}{\,2^{n+2}\pi^{(n-3)/2}\Gamma(\fr{n-1}{2})\,}
   \(\FR{\ECM}{2M_W^{}}\)^{\!n-2} \!
   \begin{pmatrix}2&0 \\ 0 &-1\end{pmatrix} ,
 \end{align}
 and extract the maximal eigenvalue,
 \bge
 \label{eq:a0-max}
   \big|a_0^{\max}\big| ~=~
   \FR{\tilde g^2}{\,2^{n+1}\pi^{(n-3)/2}
   \Gamma(\fr{n-1}{2})\,}\(\FR{\ECM}{2M_W}\)^{\!n-2} \,,
 \ede
 for which we may impose the unitarity condition (\ref{eq:UC-PW}),
 $\,\big|a_0^{\max}\big| \leqq 1\,$  (or
 $\,\big|\Re\mathfrak{e}a_0^{\max}\big| \leqq \fr{1}{2}\,$).\,
 The same amplitude (\ref{eq:a0-max})
 can also be inferred from computing the scattering of
 spin-0 and gauge-singlet state,
 $\,|0\ra = \fr{1}{\sqrt{6}\,}\(2|W_L^+W_L^-\ra +|Z_L^0Z_L^0\ra\)\,$.\,

 We show the unitarity constraint in Fig.\,\ref{Fig-PartWave}(a)-(b)
 for our HLSM-SDR, under the ansatz\footnote{We have checked
 other possible variations of (\ref{eq:DFansatz2})
 and found that our main physics picture remains.}\,
 (\ref{eq:DFansatz2}) for the dimensional flow with index
 $\,\gamma =2\,$ [plot-(a)] and $\,\gamma =1.5\,$ [plot-(b)].
 In each plot of Fig.\,\ref{Fig-PartWave}, we have also varied the transition scale
 $\,\cut =4,\,5,\,6\,$TeV, in red, purple and blue curves (from bottom to top), respectively.
 The shaded region in yellow is excluded by the unitarity bound
 $\,\big|a_0^{\max}\big| \leqq 1\,$.\,
 From this plot we see that the partial wave
 always has a rather broad ``lump" around $1.5-5$\,TeV
 and then falls off rapidly, exhibiting unitary high energy behaviors.
 For comparisons in the same plot,
 we also display the results of (i) the 4d SM with a {600}\,GeV Higgs boson, and (ii)
 the naive 4d Higgsless SM which breaks the unitarity at $\,\ECM\simeq 1.74\,$TeV.

 \begin{figure}[t]
   \begin{center}
     \includegraphics[width=0.7\textwidth]{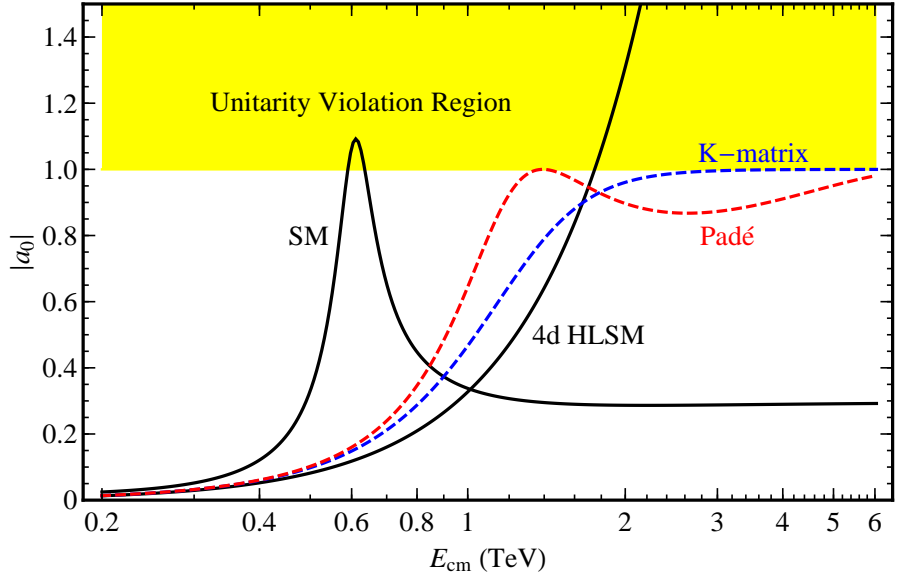}
   \end{center}
   \vspace*{-5mm}
   \caption{Same as Fig.\,\ref{Fig-PartWave}, but for the 4d Higgsless SM
   under the K-matrix and Pad\'e unitarizations in blue and red dashed curves, as comparisons
   with our HLSM-SDR. The $s$-wave amplitudes for the non-unitarized minimal 4d\,HLSM
   and for the conventional 4d SM (with a {600}\,GeV Higgs boson) are also
   shown by the black curves as the reference.
   }
   \label{fig:sWave-PadeK}
 \end{figure}

 We stress that our new mechanism of maintaining unitarity
 is via the {\it universal phase-space reduction of final states under the SDR.}
 This fully differs from the conventional 4d SM, which ensures the unitarity
 of $WW$ scattering via exchange of Higgs resonance.
 In passing, it has also been known before\,\cite{Dicus:2004rg}
 that changing the phase space can strongly affect the unitarity bound.
 As shown in Ref.\,\cite{Dicus:2004rg},
 {\it enlarged phase space} of $\,2\!\to\! {N}$\, scattering
 by proper increase of the number ${N}$ of
 final state gauge bosons can enhance the cross section and
 lead to a new class of tight unitarity limits for all light
 SM fermions (including Majorana neutrinos).
 It is interesting that our present study points to the other way around:
 {\it reducing the final-state phase space} in $\,2\!\to\! 2$\, scattering
 by lowering the spacetime dimension $n$ can significantly
 reduce the partial wave amplitudes and cross sections,
 and thus nicely restore the unitarity.
 We further note that, since the reduction of final-state phase space is caused by SDR,
 this mechanism is universal to all kinds of final-state particles, including
 fermions and vector bosons (such as transverse polarizations of
 weak gauge bosons or photons).

 For comparison with the conventional unitarizations,
 we also analyze the coupled-channel $s$-wave amplitudes for
 longitudinal scattering in the 4d Higgsless SM,
 by using the traditional K-matrix method and Pad\'e method \cite{K-Pade},
 though they are somewhat arbitrary without knowing the actual UV dynamics.
 This means that we compute the partial waves to the order of
 $\order{\ECM^4}$, and make the following resummations
 for the $\,\order{\ECM^2}\,$ and $\,\order{\ECM^4}\,$ partial waves
 $\,a^{(2)}\,$ and $\,a^{(4)}\,$,
 \beqs
 \begin{align}
    a^{}_K(\ECM) ~=~& \FR{a^{(2)}+\RE a^{(4)}}{~1-\ii(a^{(2)}+\RE a^{(4)})~} \,,
   &&  \text{(K-matrix)}\,,
   \\[1.5mm]
   a^{}_P(\ECM) ~=~& \FR{a^{(2)}}{~1-(a^{(4)}/a^{(2)})~} \,,
   &&  \text{(Pad\'e)}\,.
 \end{align}
 \eeqs
 Following the traditional K-matrix and Pad\'e procedures\,\cite{K-Pade},
 we show the corresponding $s$-wave amplitudes for the 4d Higgsless SM
 in Fig.\,\ref{fig:sWave-PadeK}.
 This should be compared with our HLSM-DSR prediction in Fig.\,\ref{Fig-PartWave}(a)-(b).

\vspace*{3mm}
\subsection{Analysis of $\,{WW}$\, Scattering Cross Sections under SDR}
\vspace*{1.5mm}

 \begin{figure*}
   \begin{center}
     \includegraphics[height=9cm,width=0.7\textwidth]{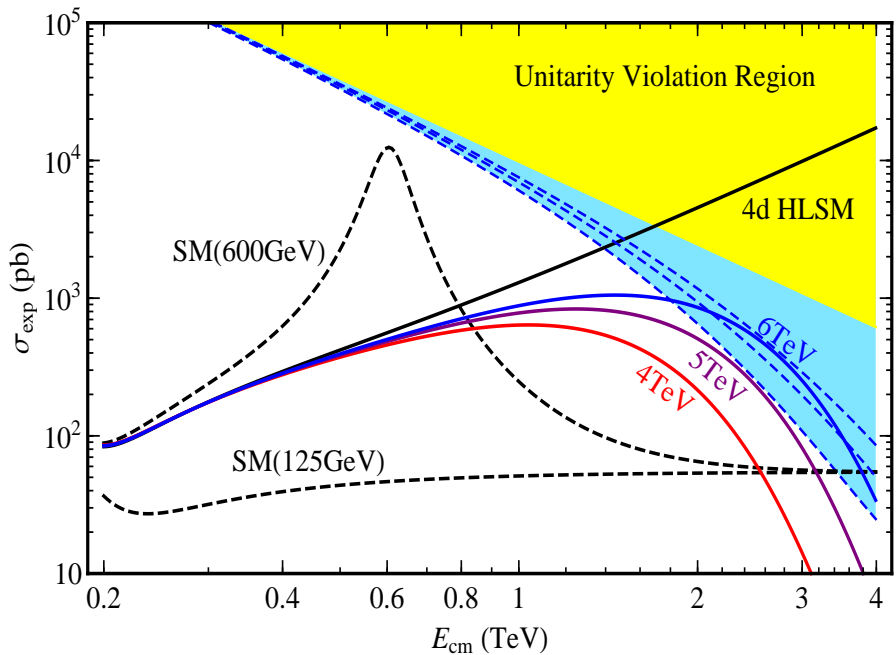}
     \vspace*{-5mm}
   \end{center}
   \caption{Cross section of $\,W_L^+W_L^-\to Z_L^0Z_L^0\,$
   as a function of the c.m.\ energy $\ECM$.
   The cross sections for the HLSM-SDR are shown
   by red, purple, and blue curves, corresponding to the UV cutoff
   $\,\cut =4,\,5,\,6\,$TeV, respectively.
   For comparisons, the cross sections for the conventional 4d SM
   with a light Higgs boson ($M_h=125$\,GeV) and a heavy Higgs boson ($M_h=600$\,GeV)
   are shown in black dashed-curves; the cross section for the 4d Higgsless SM is
   depicted by the black solid-curve.
   The shaded region with yellow color represents unitarity violation
   in 4d. The light-blue region (plus the yellow region) denotes unitarity violation
   in the HLSM-SDR, and the three blue dashed-lines, from
   bottom to top, show the unitarity bounds with
   $\,\cut =4,\,5,\,6\,$TeV, respectively.}
   \label{fig:cs-wwzz}
\end{figure*}

  In this subsection, we further analyze the $WW$ scattering sections for the HLSM-SDR.
  We study the scattering processes:
  (a).\ $W^+_LW^-_L\to Z_L^0Z_L^0$,\,
  (b).\ $W^\pm_LZ_L^0\to W^\pm_LZ_L^0$,\,
  (c).\ $W^\pm_LW^\pm_L\to W^\pm_LW^\pm_L$,\,
  which will be measured at the LHC.
  In particular, the scattering (a) contains an $s$-channel Higgs boson resonance
  for the conventional SM, while the reactions (b) and (c) do not.\footnote{%
  The scattering (b) is also expected to be sensitive to $s$-channel charged new vector bosons
  such as the $W_1^\pm$ in the extended gauge sector\,\cite{Abe:2012fb}
  and in the extra dimensional theories and its deconstruction\,\cite{He:2007ge},
  or techni-rho $\rho_{T}^\pm$ in technicolor models\,\cite{strong}.
  The scattering (c) can be used to probe the double-charged Higgs particles in some extensions
  of the SM \cite{LR}, or the non-resonance scenario \cite{nonR}, or the anomalous
  Higgs-gauge couplings of a light Higgs boson \cite{He:2002qi,HVV-2}.}\,
  However, the new predictions of our HLSM-SDR arise from the spontaneous
  reduction of spacetime dimensions at high energies due to the nonperturbative dynamics
  of quantum gravity, so {\it they are universal and show up in all $WW$ scattering channels.}
  This is an essential feature of our HLSM-SDR and will play an important role in discriminating
  the HLSM-SDR from all other models of the electroweak symmetry breaking.

  To evaluate cross sections in a spacetime with varying dimensions for our HLSM-SDR,
  we pay special attention to the phase space integral.
  For clarity, we present the systematical derivations in Appendix-A.
  In $4$-dimensions, the unitarity bounds for cross sections are
  given by \cite{Dicus:2004rg},
 \beqa
   \si_\el^{} ~\leqq~\FR{16\pi\rh_e^{}}{\ECM^2} \,, &~~~&
   \si_\inel^{}~\leqq~\FR{4\pi\rh_e^{}}{\ECM^2} \,.
 \eeqa

 In general $n$-dimensions, we have derived the formulas
 in (\ref{eq:CSuni-n}) of Appendix-A,
 and find that the above bound should be extended to,
 \beqa
   \label{UniConCS}
   \si_\el ~\leqq~ \FR{\lam_n\rh_e^{}}{\,N_0^\nu\ECM^{n-2}\,} \,,
   &~~~&
   \si_\inel ~\leqq~ \FR{\lam_n\rh_e^{}}{\,4N_0^\nu\ECM^{n-2}\,} \,,
 \eeqa
 where $\,\lam_n^{}\,$ and $\,N_0^\nu\,$ are defined in (\ref{eq:la-nu-Nl}).

\begin{figure*}
   \begin{center}
     \includegraphics[width=0.7\textwidth]{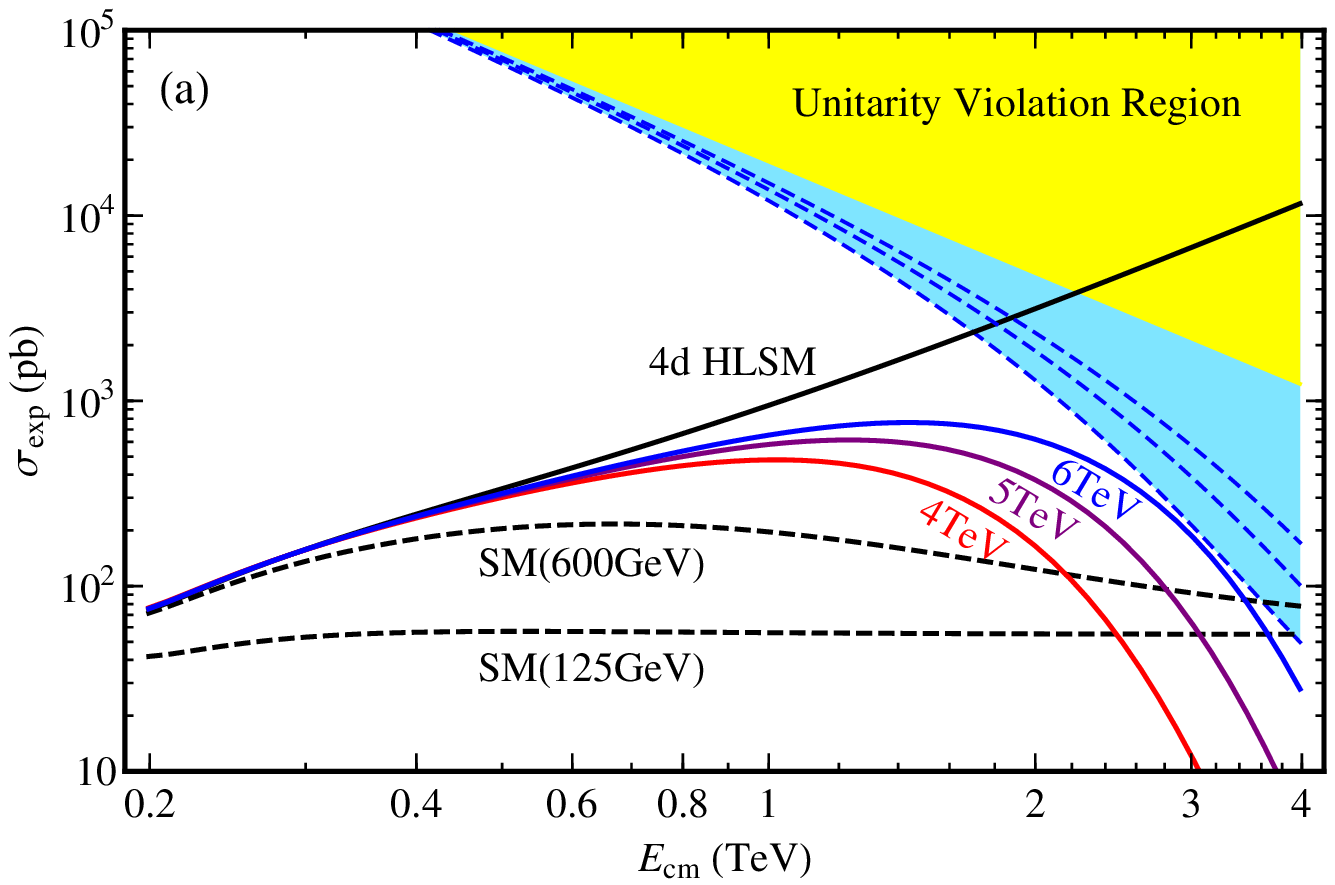} \\
     \includegraphics[width=0.7\textwidth]{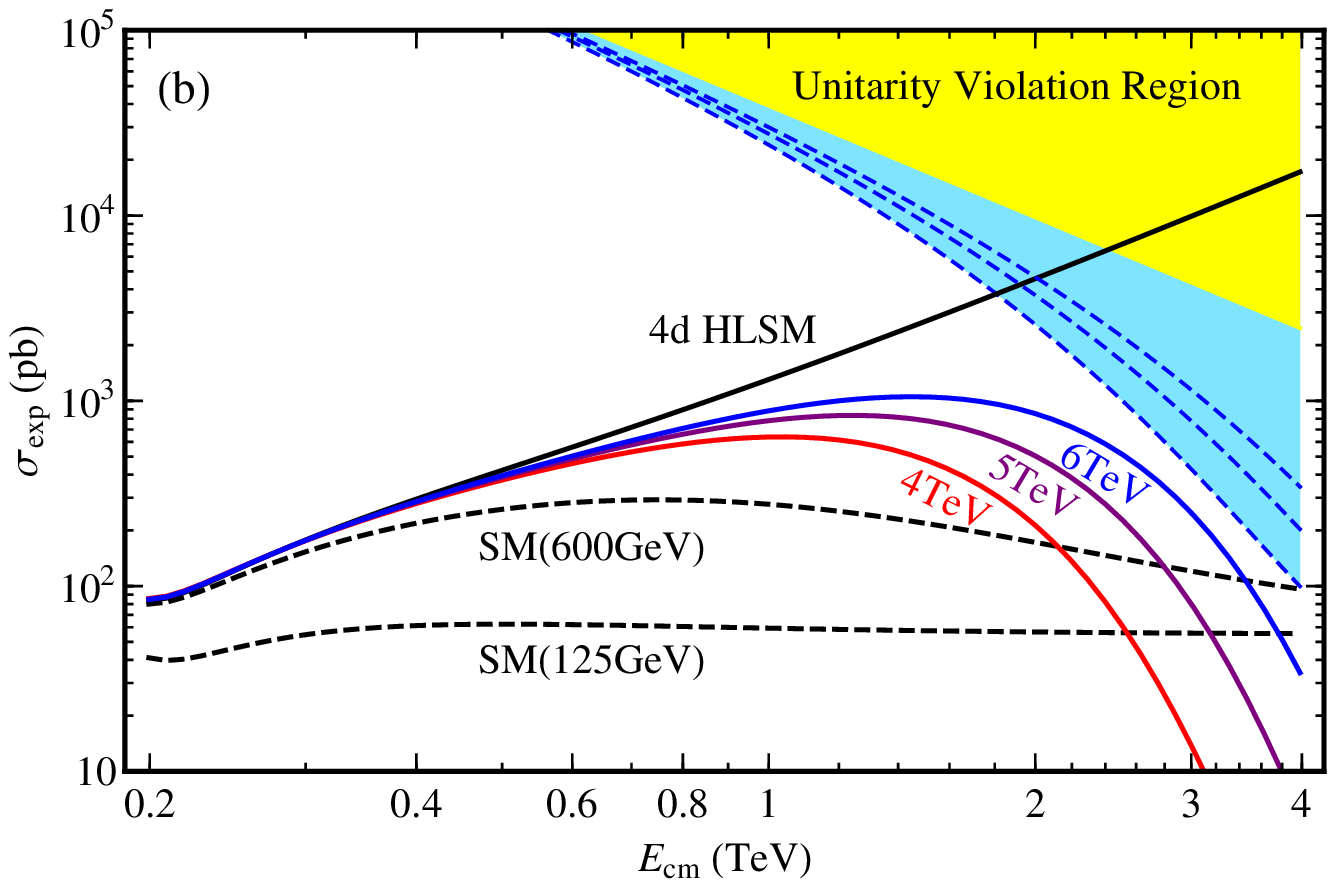}
   \end{center}
   \vspace*{-5mm}
   \caption{Plot-(a) depicts the cross section of $\,W_L^\pm Z_L^0\to W_L^\pm Z_L^0\,$
   as a function of the c.m.\ energy $\ECM$, and plot-(b) shows
   the cross section of $\,W_L^\pm W_L^\pm\to W_L^\pm W_L^\pm\,$
   as a function of $\ECM$.
   In each plot, the cross sections for the HLSM-SDR are shown
   by red, purple, and blue curves, corresponding to the UV cutoff
   $\,\cut =4,\,5,\,6\,$TeV, respectively.
   For comparisons, the cross sections for the conventional 4d SM
   with a light Higgs boson ($M_h=125$\,GeV) and a heavy Higgs boson ($M_h=600$\,GeV)
   are shown in black dashed-curves; the cross section for the 4d Higgsless SM is
   depicted by the black solid-curve.
   The shaded region with yellow color represents unitarity violation
   in 4d. The light-blue region (plus the yellow region) denotes unitarity violation
   in the HLSM-SDR, and the three blue dashed-lines, from
   bottom to top, show the unitarity bounds with $\,\cut =4,\,5,\,6\,$TeV, respectively.
   }
   \label{fig:cs-wz-w+w+}
 \end{figure*}

  With these, we present our predictions for the HLSM-SDR
  in Fig.\,\ref{fig:cs-wwzz} and Fig.\,\ref{fig:cs-wz-w+w+} for $\,\gamma =1.5\,$.\,
  As comparisons, we have computed the cross sections of these processes for the
  conventional SM with a light Higgs boson ($M_h=125$\,GeV)
  and a heavy Higgs boson ($M_h=600$\,GeV), depicted by the black dashed-curves.
  We also evaluate the cross section for the 4d Higgsless SM as shown by
  the black solid-curve.
  In each plot, we show our analysis for different transition scales,
  $\,\cut =4,\,5,\,6$\,TeV, respectively.
  The unitarity bound for the cross section is also depicted in
  Figs.\,\ref{fig:cs-wwzz}-\ref{fig:cs-wz-w+w+}.
  The regions shaded by yellow and light-blue colors represent unitarity violation
  in 4d and in the HLSM-SDR, respectively, and the three blue dashed-curves,
  from bottom to top, show the unitarity bounds with
  $\,\cut =4,\,5,\,6\,$TeV, respectively.
  From these, we see how the SDR works as a mechanism
  to unitarize the high energy behaviors
  of cross sections without invoking extra particles (such as the SM Higgs boson).
  This can be discriminated from other unitarization schemes of $\,WW$ scattering
  at the LHC.

\begin{figure*}
   \begin{center}
     \includegraphics[width=0.7\textwidth]{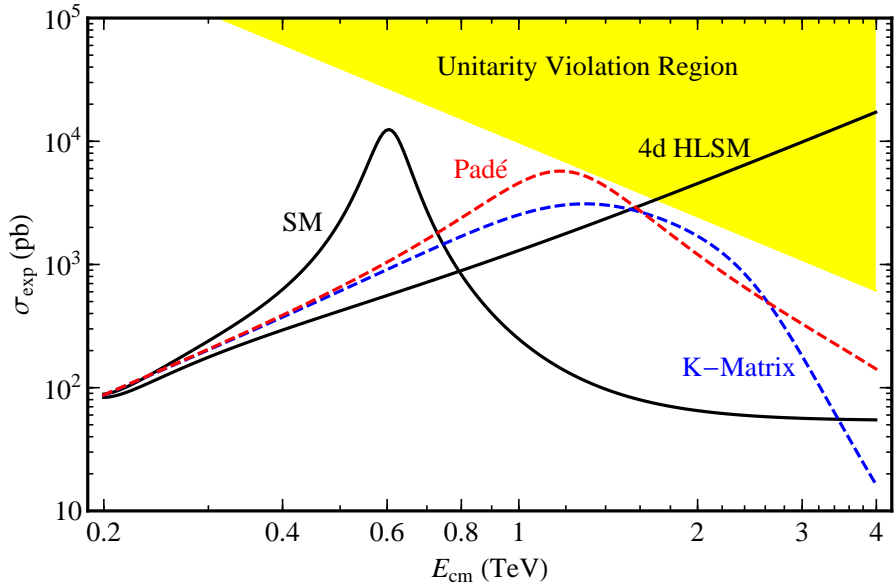}
     \vspace*{-5mm}
   \end{center}
   \caption{Same as Fig.\,\ref{fig:cs-wwzz} for the cross section of
   $\,W_L^+W_L^-\to Z_L^0Z_L^0\,$,\, but we we show it in the 4d Higgsless SM
   under the K-matrix (red-dashed) and Pad\'e (blue-dashed) unitarizations,
   as comparisons with our HLSM-SDR. The shaded area with yellow color
   is the unitarity violation region.
   The cross section for the conventional 4d SM with a {600}\,GeV Higgs boson
   is also shown as a reference.}
   \label{fig:cs-wwzz-kp}
 \end{figure*}

 For comparisons with the current HLSM-SDR, we further plot in Fig.\,\ref{fig:cs-wwzz-kp}
  the cross section of $\,W_L^+W_L^-\to Z_L^0Z_L^0\,$
  in the 4d Higgsless SM under the K-matrix (red-dashed) and Pad\'e (blue-dashed)
  unitarizations. The shaded area with yellow color denotes the unitarity violation region.
  The cross section for the conventional 4d SM with a {600}\,GeV Higgs boson
  is also shown as a reference.  In passing, we would like to note that
  the unitarity of $WW$ scattering in generic 4d technicolor-type theories
  was recently studied by Sannino and collaborators \cite{Sannino}.

\vspace*{2mm}

 Finally, we clarify how to compare the SDR cross section computed above
 with what will be measured by the experiments.
 Note that the cross section $\,\si\,$ in spacetime dimensions other than $4$ also
 has different mass-dimensions, equal to $\,[\si]=2-n$\,.\,
 But the experimentally measured cross section $\,\si_{\text{exp}}^{}$\,
 always has mass-dimension $\,-2$\,,\, as the detectors can record events only in 4d\,.\,
 So we need to convert the theory cross section $\,\si$\,
 under the SDR to $\,\si_{\text{exp}}^{}$\,,\,
 where the extra mass-dimensions of $\,\si\,$ should be scaled
 by the involved energy scale of the reaction, namely, the c.m.\ energy $\,\ECM$\,,
 \bge
   \si_{\text{exp}} ~=~ \si \,\ECM^{n-4} \,.
 \ede
 In Figs.\,\ref{fig:cs-wwzz}-\ref{fig:cs-wz-w+w+},
 the unitarity bound with SDR (the blue region)
 has also been rescaled by $\,\ECM^{n-4}$\,
 to keep its mass-dimension in accord with experimental measurements.

 In passing,
 we also note that our present unitarity analysis only depends on the total cross sections
 where the possible azimuthal angles are formally integrated out, as shown in
 (\ref{eq:dp^n-1})-(\ref{PhSpaceInt3}),
 by the same way as the conventional dimensional regularization method for doing
 momentum-space integrals in $\,n<4$\, dimensions.
 The dependence of differential cross sections on the azimuthal angle
 in a theory with SDR relies on the detailed geometrical structure of the model.
 But this is irrelevant to our current unitarity study, and will be considered
 in future works.

\vspace*{3mm}
\section{Weak Boson Scattering in Higgsful SM with SDR}
\vspace*{1.5mm}

As demonstrated in the previous section,
the SDR can play a unique role to unitarize the weak boson scattering
at TeV scales in the absence of Higgs boson. We note that
the SDR can equally occur at TeV scales in the presence of a light Higgs boson.
Especially, such quantum-gravity-induced SDR at TeV scales provides a
natural solution to the hierarchy problem (fine-tuning problem) \cite{unnatural}
that plagues the Higgs boson in the conventional 4d SM.
(In passing, we also note that some interesting phenomenological studies\,\cite{litim}
on TeV-scale quantum gravity appeared in a very different context
under the asymptotical safety scenario.)

For the TeV-scale SDR, new physics effects induced by the quantum gravity
are expected in the low energy effective theory.
Thus, the Higgs boson can behave as non-SM-like and encode such new physics in
its anomalous couplings with $WW$ and $ZZ$ gauge bosons, as well as in its self-couplings.
In the conventional 4d SM with a non-SM Higgs boson\,\cite{HVV-0},
it was found that the $WW$ scattering has non-canceled $E^2$
behavior\,\cite{He:2002qi} in the TeV range, and can be probed
at the LHC \cite{He:2002qi,HVV-2}.  But additional new physics is required
to unitarize the non-canceled $E^2$ contributions to the $WW$ scattering.
In this section, we study the unitarity of weak boson scattering
in the Higgsful SM with SDR (HFSM-SDR),
which contains a light Higgs boson of mass 125\,GeV, but with anomalous gauge couplings.
We will demonstrate that under the SDR,
the corresponding $WW$ scattering cross sections become
unitary at TeV scale, but exhibit different behaviors.

Let us start from the nonlinear realization of the SM Higgs boson $\,h\,$ in terms of
the $2\!\times\!2$ matrix field $\,\Phi\,$ as defined in (\ref{eq:Phi-Sigma}),
where $\,h\,$ is a gauge-singlet. From (\ref{eq:L-Sigma}), we inspect
the general effective Lagrangian for the Higgs sector
(up to dimension-4 operators in 4d) \cite{HVV-0}\cite{He:2002qi},
 \beqa
 \label{eq:L-h-Sigma}
 \ld_H^{} &\,=\,&
 \frac{1}{4}\(v^2+2\ka v h +\ka' h^2\)
 \tr\!\!\left[(\D^\mu\Sigma)^\dag(\D_\mu\Sigma)\right]
 \n\\[1mm]
 &&
 +\frac{1}{2}\dif_\mu^{}h\dif^\mu h
 -\frac{1}{2}M_h^2h^2 - \frac{\la_3^{}}{3!}vh^3
 +\frac{\la_4^{}}{4!}h^4 \,,~~~~~
 \eeqa
where $\,\De\ka \equiv\kappa-1\,$ and $\,\De\ka' \equiv\kappa'-1\,$ represent
the anomalous gauge couplings of the Higgs boson $h$ with $WW$ and $ZZ$.\,
For cubic and quartic Higgs self-couplings, the conventional 4d SM predicts
$\,\la_{3}^{}=\la_{4}^{}=\la_0^{}=3M_h^2/v^2\,$,\,
but the general effective Lagrangian
(\ref{eq:L-h-Sigma}) also allows anomalous self-couplings
$\,\la_{3}^{},\la_{4}^{}\neq \la_{0}^{}\,$.\,
In parallel, from (\ref{eq:L-FmassSigma}), we note the general effective Lagrangian
for the fermion-Higgs sector (up to dimension-4 operators in 4d),
 \beqa
 \label{eq:L-FH}
 \ld_{FH}^{}
 ~=~ - \overline{\Psi_L^{}}\,\Sigma
 \!\left[M_f^{}
  +\(\frac{M_f^{}}{v} + \hf\De{\cal Y}_f^{}\)h\,\right]\!\Psi_R^{}
  + \text{h.c.}\,,
 \eeqa
where  $\,\De {\cal Y}_f^{}=\text{diag}(\De y_1^{},\,\De y_2^{})\,$ denotes the
anomalous Yukawa couplings of the non-standard Higgs boson.
In the current analysis,
we will focus on the anomalous gauge coupling  $\,\De\ka\neq 0$\,
in (\ref{eq:L-h-Sigma}) for the study of weak boson scattering.
Under unitary gauge, we can infer from (\ref{eq:L-h-Sigma}) the following
anomalous cubic and quartic gauge-Higgs interactions,
\beqa
\label{eq:dL-HVV}
 \De\ld_H^{} ~=~
 \left(\,\De\ka\, vh + \fr{1}{2}\De\ka'\, h^2\,\right)\!
 \left[ \frac{\,2M_W^2\,}{v^2}W^+_\mu W^{-\mu}
       +\frac{\,M_Z^2\,}{v^2}Z_\mu Z^\mu \right] .
\eeqa
For our present construction of the Higgsful SM with TeV-scale SDR, the new physics
effects will be induced by the non-perturbative dynamics of quantum gravity above
the transition scale $\cut$,\, and can be parameterized via leading anomalous couplings
in the effective Lagrangian, as shown above.

We note that for the conventional 4d SM,
besides the hierarchy problem\,\cite{unnatural}, it further suffers
constraints from the Higgs vacuum instability\,\cite{VacS}
and the triviality of Higgs self-interactions\,\cite{trivial}.
For such a 4d SM to be valid up to Planck scale, it is found that the SM Higgs boson mass
should be bounded from below\,\cite{VacS} and from above\,\cite{trivial2Loop},
$\,133\,\text{GeV} \lesssim M_h^{} \lesssim 180\,$GeV \cite{EW-rev}.
This means that any Higgs mass outside this window will indicate a non-standard Higgs boson
in association with {\it new physics.}
The Higgs boson in our present model under the TeV-scale SDR has
anomalous couplings induced from quantum gravity and thus should belong to this case.
For the numerical analysis below, we will consider
$\,M_h^{}=125$\,GeV for a light non-standard Higgs boson, based on the
current Higgs boson direct searches at the LHC\,\cite{LHCnew}.

The recent LHC data\,\cite{LHCnew} can already place
interesting constraints on the anomalous cubic Higgs-gauge-coupling
$\Delta\kappa$ in (\ref{eq:dL-HVV}).
Using the latest model-independent fitting result of \cite{Hfit}, we find the
$\Delta\kappa$ is bounded into the range,
$\,-0.32 < \Delta\kappa < 0.25\,$ at $2\sigma$ level, or,
$\,-0.46 < \Delta\kappa < 0.39\,$ at $3\sigma$ level.

In the high energy region with $\,E_{\text{cm}}^2\gg M_W^2,M_h^2\,$,\,
we note that the anomalous cubic interactions in (\ref{eq:dL-HVV})
induce non-canceled $\ECM^2$ terms in the longitudinal gauge boson
scattering amplitudes. For the processes (\ref{SingProc1})-(\ref{SingProc2}),
we find the non-canceled $\order{\ECM^2}$ amplitudes,
 \beqs
 \label{eq:T-4V-AC}
 \beqa
 \label{eq:WWWW-ano}
   \mathcal{T}[W_L^+W_L^-\to W_L^+W_L^-] &=&
   (1\!-\!\ka^2)\FR{g^2\ECM^2}{\,8M_W^2\,}(1+\cos\theta) + \order{\ECM^0} \,,
\\[1.5mm]
\label{eq:WWZZ-ano}
  \mathcal{T}[W_L^+W_L^-\to \fr{1}{\sqrt{2}\,}Z_L^0Z_L^0] &=&
   (1\!-\!\ka^2)\FR{g^2\ECM^2}{\,4\sqrt{2}M_W^2\,} + \order{\ECM^0} \,.
 \eeqa
 \eeqs
 They are expected to eventually violate the unitarity as the increase of scattering
 energy $\,\ECM\,$ in the conventional 4d formulation.
 But, for our present SDR formulation, these scattering amplitudes
 and the corresponding cross sections will be unitarized
 at TeV scales in a rather universal way,
 due to the spontaneous dimensional reduction.

 In parallel to Sec.\,3.1, we first study the unitarity constraints on
 the anomalous Higgs coupling $\,\De\ka\,$ via the coupled channel analysis
 of partial wave amplitudes of weak boson scattering.
 Thus, from (\ref{eq:WWWW-ano}), we derive a
 $2\times 2$ matrix of the $s$-wave amplitudes in $n$-dimensions
 for the initial/final states
 $\,|W_L^+W_L^-\ra\,$ and $\,\fr{1}{\sqrt{2}\,}|Z_L^0Z_L^0\ra\,$,\,
 \begin{align}
 \label{eq:a0-coup-dk}
   a_{0}^{\text{coup}} ~=~
   \FR{(1\!-\!\ka^2)\,\tilde g^2}{\,2^{n+2}_{}\pi^{\fr{n-3}{2}}
   \Gamma(\fr{n-1}{2})\,}\(\FR{\ECM}{2M_W^{}}\)^{\!n-2}\!
   \begin{pmatrix}
   1 & \,\sqrt{2}\,
   \\[1.5mm]
   \sqrt{2} & 0
   \end{pmatrix},
 \end{align}
 where we made use of (\ref{eq:PW-nDim}) and
 $\,\tilde{g}\,$ is defined by (\ref{eq:gt}).
 From diagonalizing the matrix (\ref{eq:a0-coup-dk}), we extract the
 maximal eigenvalue amplitude,
 \bge
 \label{eq:a0-max-dk}
   \big|a_0^{\max}\big| ~=~
   \FR{(1\!-\!\ka^2)\,\tilde g^2}{\,2^{n+1}\pi^{(n-3)/2}
   \Gamma(\fr{n-1}{2})\,}\(\FR{\ECM}{2M_W}\)^{\!n-2}
   \equiv~ (1\!-\!\ka^2) {\cal A}_n^{}(\ECM)\,.
 \ede
 With this, we will impose the unitarity condition (\ref{eq:UC-PW}),
 $\,\big|a_0^{\max}\big| \leqq 1\,$,\, and arrive at
\beqa
\label{eq:uni-sol-dk}
\sqrt{1-\A_n^{-1}} - 1 ~\leqq~ \De\ka ~\leqq~
\sqrt{1+\A_n^{-1}} - 1 \,,
\eeqa
where $\,\A_n^{}\,$ is defined in (\ref{eq:a0-max-dk}), and
the lower bound exists only if $\,\A_n^{}\geqq 1\,$.\,
In (\ref{eq:uni-sol-dk}),
we have considered the anomalous coupling within the range,
$\,|\De\ka| < 1\,$,\,
as indicated by the current fitting\,\cite{Hfit} to the LHC data,
$\,-0.46 < \Delta\kappa < 0.39\,$ ($3\sigma$ level).

\begin{figure*}[t]
   \begin{center}
     \includegraphics[width=0.7\textwidth]{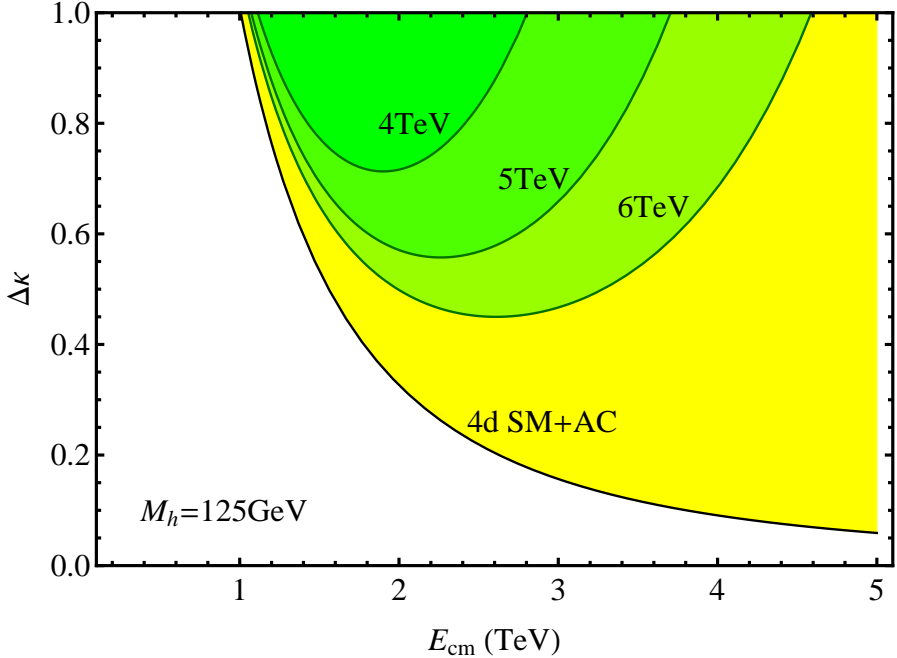}
     \vspace*{-5mm}
   \end{center}
   \caption{Unitarity constraint on the anomalous Higgs coupling $\,\De\ka$\,
   as a function of c.m.\ energy $\ECM$ of weak boson scattering, via coupled
   channel analysis of partial waves. The curves from top to bottom correspond to
   the unitarity limits of $\,\cut =4,\,5,\,6\,$TeV in the HFSM-SDR and
   to the limit in the 4d SM with the same anomalous coupling $\,\De\ka\,$
   (labeled by ``4d SM+AC"), respectively. The shaded region above each curve
   is excluded by the unitarity bound.
   }
   \label{fig:uni-dk}
 \end{figure*}

In Fig.\,\ref{fig:uni-dk}, we present the unitarity constraints
on the anomalous coupling $\,\De\ka\,$ as a function of the
scattering energy $\ECM$.
The three upper curves (from top to bottom) correspond to
the unitarity bounds of $\,\cut =4,\,5,\,6\,$TeV in the HFSM-SDR
($\gamma = 1.5$), respectively;
while the lowest curve (labeled by ``4d SM+AC") gives
the bound in the 4d SM with the anomalous coupling
$\,\De\ka\,$.\, The shaded region above each curve
is excluded by the unitarity limit.
We see that the unitarity bounds on $\,\De\ka\,$ of the HFSM-SDR
are much weaker than that in the usual 4d SM.
In the numerical analysis, we have used both the expanded amplitude
(\ref{eq:a0-max-dk}) and the exact amplitude.
We find that the expanded result (\ref{eq:a0-max-dk}) holds well and thus
the constraints in Fig.\,\ref{fig:uni-dk} are insensitive to the Higgs mass.
Furthermore, inspecting $\,\A_n^{}(\ECM)\,$ in (\ref{eq:a0-max-dk})
we find that $\,\A_n^{} < 1\,$ always holds, so the lower bound
of (\ref{eq:uni-sol-dk}) is absent for the HFSM-SDR. But
the 4d SM with nonzero $\,\De\ka\,$ will suffer both the lower
and upper limits of (\ref{eq:uni-sol-dk}) since for $\,n=4\,$
we have $\,\A_4^{} = E_{\text{cm}}^2/(16\pi v^2)\,$ and
$\,\A_4^{}> 1\,$ holds for $\,\ECM > 1.74\,$TeV.
For the current purpose of comparison with the unitarity bounds
on the HFSM-SDR, we do not show the lower bound for the 4d SM
in Fig.\,\ref{fig:uni-dk}.

\begin{figure*}
   \begin{center}
     \includegraphics[width=0.7\textwidth]{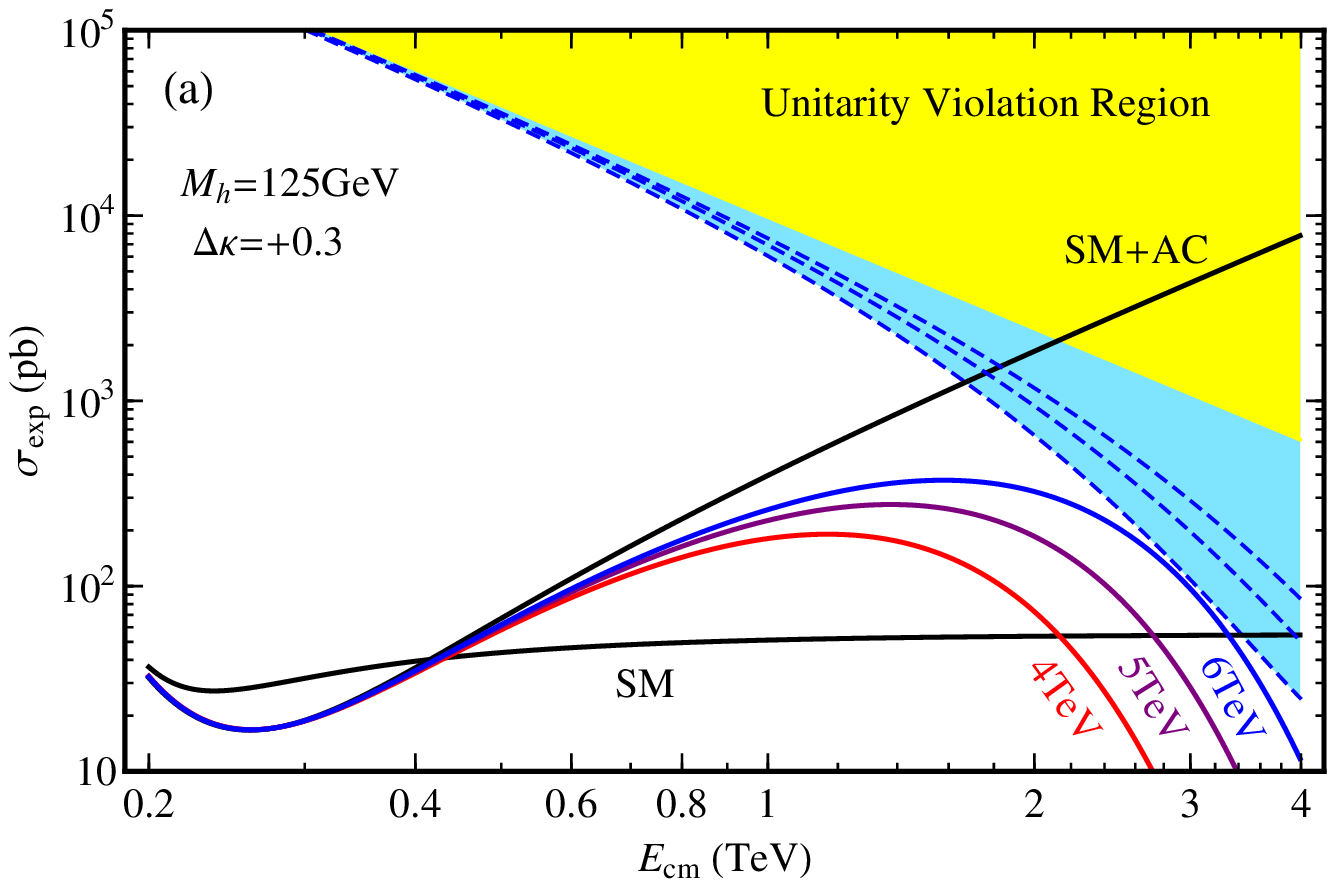}\\
     \includegraphics[width=0.7\textwidth]{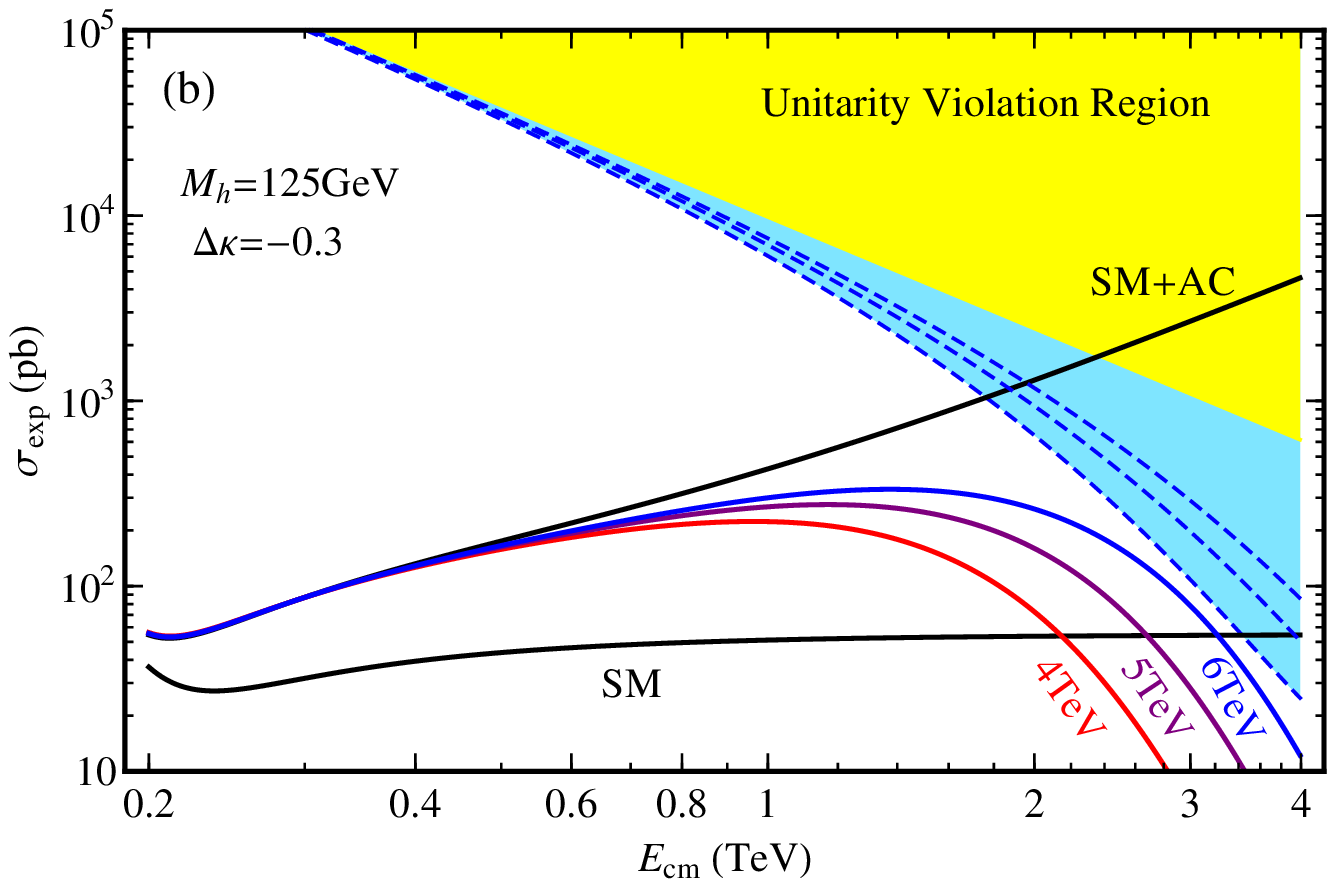}
     \vspace*{-5mm}
   \end{center}
   \caption{Cross section of $\,W_L^+W_L^-\to Z_L^0Z_L^0\,$
   as a function of c.m.\ energy $\ECM$.   In each plot,
   the cross sections for the HFSM-SDR with nonzero anomalous coupling
   $\,\De\ka\,$ are shown by red, purple, and blue curves,
   corresponding to the UV cutoff scale
   $\,\cut =4,\,5,\,6\,$TeV, respectively.
   For comparisons, the cross sections for the conventional 4d-SM
   with $\,M_h=125$\,GeV (labeled by ``SM") and the 4d-SM with the
   same anomalous coupling (labeled by ``SM+AC") are shown by the black curves.
   The shaded area with yellow color denotes the unitarity violation
   region in 4d, while the light-blue shaded area (plus the yellow region) shows the
   unitarity violation region for the HFSM-SDR, where the blue dashed-curves,
   from bottom to top, display the corresponding upper unitarity bounds for
   $\,\cut =4,\,5,\,6\,$TeV, respectively.}
   \label{fig:cs-wwzz-ac-125}
 \end{figure*}

\begin{figure*}
   \begin{center}
     \includegraphics[width=0.7\textwidth]{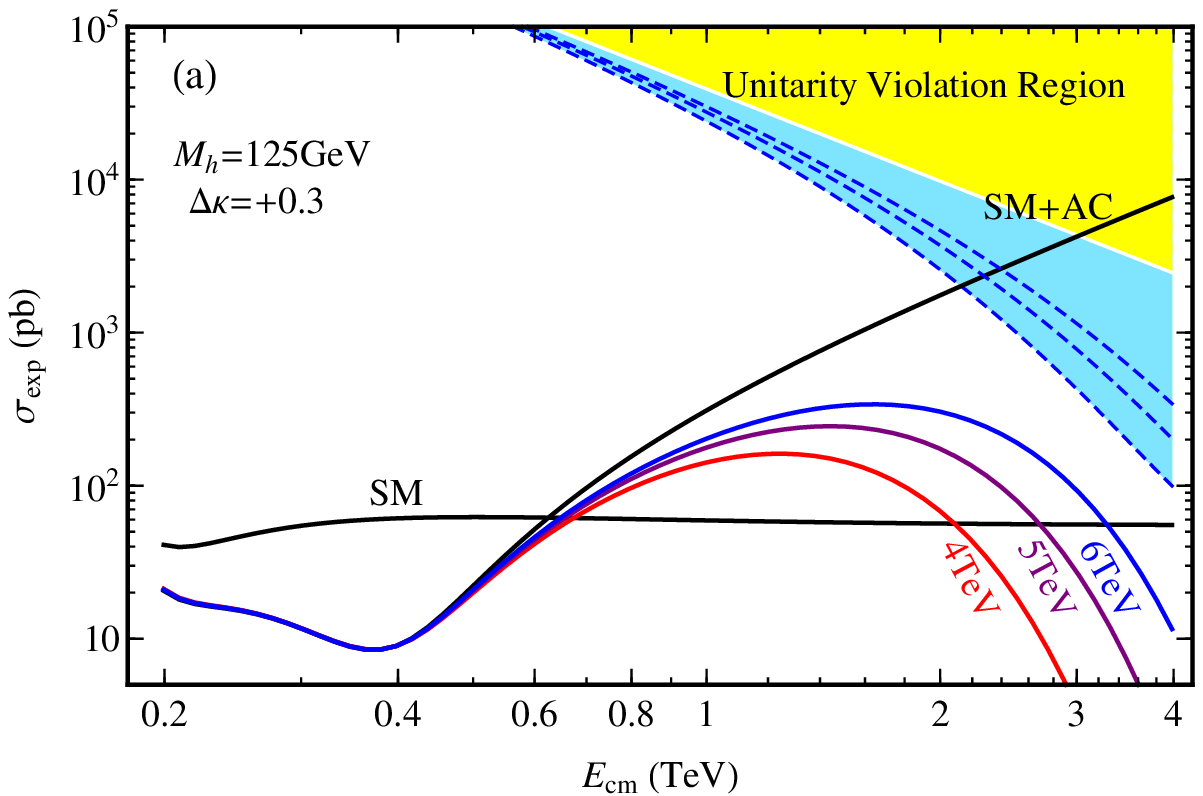}\\
     \includegraphics[width=0.7\textwidth]{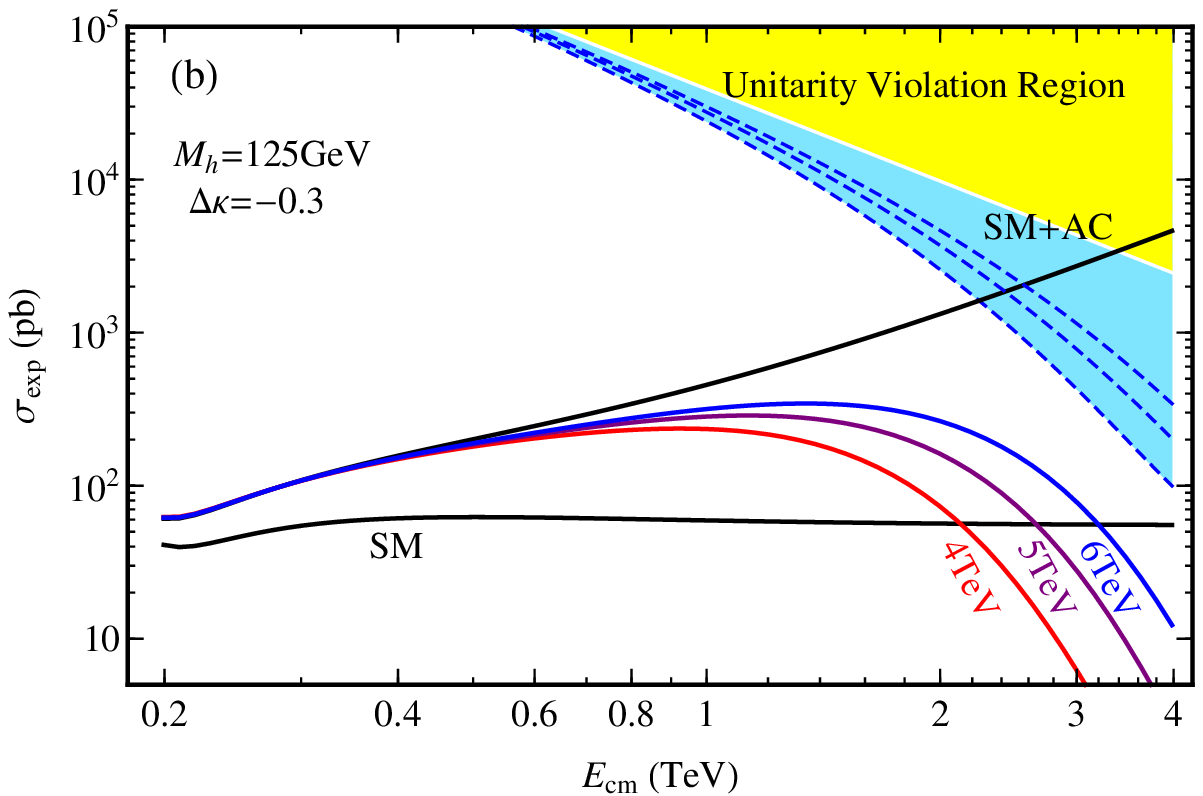}
     \vspace*{-5mm}
   \end{center}
   \caption{Cross section of $\,W_L^\pm W_L^\pm \to W_L^\pm W_L^\pm\,$
   as a function of c.m.\ energy $\ECM$. In each plot,
   the cross sections for the HFSM-SDR with nonzero anomalous coupling
   $\,\De\ka\,$ are shown by red, purple, and blue curves,
   corresponding to the UV cutoff scale
   $\,\cut =4,\,5,\,6\,$TeV, respectively.
   For comparisons, the cross sections for the conventional 4d-SM
   with $\,M_h=125$\,GeV (labeled by ``SM") and the 4d-SM with the
   same anomalous coupling (labeled by ``SM+AC") are shown by the black curves.
   The shaded area with yellow color denotes the unitarity violation
   region in 4d, while the light-blue area (plus the yellow region) shows the
   unitarity violation region for the HFSM-SDR, where the blue dashed-curves,
   from bottom to top, display the corresponding upper unitarity bounds for
   $\,\cut =4,\,5,\,6\,$TeV, respectively.
   }
   \label{fig:cs-wpwp-ac-125}
 \end{figure*}

 Next, we study the cross section of the scattering process
 $\,W_L^+W_L^-\to Z_L^0Z_L^0\,$.\,
 In Fig.\,\ref{fig:cs-wwzz-ac-125}(a)-(b)
 we show the scattering cross section for this channel,
 where we consider a light Higgs boson with the sample mass $\,M_h=125\,$GeV
 and anomalous couplings $\,\De\ka =\pm 0.3\,$.\,
 Here our choice of $\,\De\ka =\pm 0.3\,$ is consistent with the
 unitarity constraints in Fig.\,\ref{fig:uni-dk} and the model-independent
 fits to the LHC data \cite{Hfit}.
 The red, purple and blue curves correspond to the
 SDR UV cutoff $\,\cut =4,\,5,\,6\,$TeV, respectively,
 where we set the SDR index parameter $\,\gamma=1.5\,$.\,
 The shaded area with yellow color represents the unitarity violation region in 4d.
 The shaded area with light-blue color shows the unitarity violation region for
 the HFSM-SDR, where the three blue dashed-curves, from bottom to top, display the
 corresponding (upper) unitarity limits for
 $\,\cut =4,\,5,\,6\,$TeV, respectively.
 In each plot, the upper black curve (marked by ``SM$+$AC") denotes the result of conventional
 4d SM with the same Higgs mass $\,M_h=125\,$GeV and the same anomalous coupling
 $\,\De\ka =\pm 0.3\,$.\, The lower black curve (marked by ``SM") is nearly flat and denotes
 the usual 4d-SM with the same Higgs mass but zero anomalous coupling $\,\De\ka =0\,$.\,
 This process contains an $s$-channel Higgs-exchange as in Fig.\,\ref{Fig-WWZZ}(d), but
 it does not show up in Fig.\,\ref{fig:cs-wwzz-ac-125}(a)-(b) because the light Higgs boson
 has its mass ($M_h=125\,$GeV) below the $ZZ$ threshold.
 In contrast, our SDR-unitarized cross sections have sizable excesses above the
 flat black curve of the 4d-SM with $\,M_h=125\,$GeV and $\,\De\ka =0\,$,\,
 but then fall off around the $2-4$\,TeV region,
 consistent with the corresponding unitarity limits.
 We also note that the usual non-unitarized 4d-SM with nonzero
 anomalous coupling $\,\De\ka \neq 0\,$  has its cross section monotonically increase and
 eventually violate unitarity around $\,\ECM =2.2$\,TeV for this scattering channel.
 As a final remark, we note that plot-(a) with $\,\De\ka =+0.3 > 0\,$
 shows a small valley around $\,\ECM =260$\,GeV, which falls below the flat curve of
 the pure 4d SM. This is because the anomalous amplitude is proportional to the factor
 $\,(1-\ka^2)=-2\De\ka - \De\ka^2\,$ as in (\ref{eq:T-4V-AC})
 and thus dominated by the term $\,-2\De\ka\,$,\,
 which is negative for $\,\De\ka > 0\,$ and thus cancels against to
 the squared pure-SM-amplitude in the cross section calculation. This is why such
 a valley shows up for $\,\De\ka > 0\,$ in Fig.\,\ref{fig:cs-wwzz-ac-125}(a), but
 not in Fig.\,\ref{fig:cs-wwzz-ac-125}(b) with $\,\De\ka < 0\,$ .
 (For the same reason, we will see that Fig.\,\ref{fig:cs-wpwp-ac-125}(a)
 with $\,\De\ka > 0\,$ also shows a valley below the flat curve of the pure SM,
 but Fig.\,\ref{fig:cs-wpwp-ac-125}(b) does not.)

 In Fig.\,\ref{fig:cs-wpwp-ac-125},
 we study the like-sign weak boson scattering process,
 $\,W_L^\pm W_L^\pm \to W_L^\pm W_L^\pm\,$, where we choose the same inputs
 as in Fig.\,\ref{fig:cs-wwzz-ac-125}, namely,
 $\,M_h=125\,$GeV and $\,\De\ka =\pm 0.3\,$.\,
 All the labels have the same meaning as in
 Fig.\,\ref{fig:cs-wwzz-ac-125}.
 This process only involves Higgs-exchanges in $t$ and $u$ channels.
 We see that the conventional 4d SM with $\,M_h=125\,$GeV gives only
 a rather flat cross section over the full range of scattering energy.
 This is expected since the scattering amplitude is unitarized by
 this light SM Higgs boson. But, for the 4d SM with an anomalous coupling
 $\Delta\ka\neq 0$ (denoted as ``SM+AC"),
 the cross section has power-law growth with the increase of energy
 and quickly violates unitarity around $\,\ECM \sim 3$\,TeV.
 On the other hand, our HFSM-SDR with nonzero $\De\ka$ predicts unitarized
 scattering cross sections, which exhibit distinctive behaviors from
 the 4d SM and the 4d SM+AC, over the energy range around
 $\,\ECM = 0.6-4\,$TeV.
 Plot-(a) has input a positive anomalous coupling $\,\De\ka =0.3\,$,
 this creates a large valley around $\,\ECM = 0.38\,$TeV for the cross section curve,
 as compared to the pure 4d SM prediction. Then, as energy increases to above 0.7\,TeV,
 each cross section under SDR gets unitarized and forms a broad lump around
 $\,0.7-3\,$TeV, which shows significant excess above the flat curve of the 4d SM.
 In plot-(b), we have a negative anomalous coupling
 $\,\De\ka =-0.3\,$. As a result, the signal curves are generally above that of
 the 4d SM all the way up to about $\,\ECM \simeq 2$\,TeV, where the cross sections
 start to get fully unitarized and substantially deviate from the naive 4d SM+AC.
 But, the excess in plot-(b) shows no peak-like structure, and
 thus is harder to be discriminated from the 4d SM curve.

\vspace*{3mm}

Before concluding this section, we make a clarification on
the validity range of our effective theory of the SDR.
The validity range lies between the $WW$\,($ZZ$) threshold
(about 160-180\,GeV) and the UV-cutoff $\cut =\order{5\text{TeV}}$.
For instance, our Figs.\,\ref{fig:cs-wwzz}-\ref{fig:cs-wz-w+w+}
and Fig.\,\ref{fig:cs-wwzz-ac-125}-\ref{fig:cs-wpwp-ac-125} demonstrate that
the relevant scattering energy $E_{\text{cm}}^{}$
(at which our model can be distinguished from the 4d-SM and 4d-HLSM at the LHC)
is always within $\,0.2-3$\,TeV,\, and thus is {\it significantly below
4\,TeV}\,(\,$\leqq \cut$\,).
Furthermore, for this energy region $0.2-3$\,TeV (relevant to the LHC probe),
and given the typical model-inputs ($\,\cut=5\,$TeV and $\gamma =1.5$\,),\,
we can directly infer the dimensional flow from our Eq.\,(\ref{eq:DFansatz2}),
$\,n\simeq 3.94-3.07\,$.\,
This is {\it significantly above $\,n=2\,$,}
and thus our effective theory should hold well.
Hence, our effective theory study does not depend on
any detail of the UV dynamics around the cutoff $\cut$ and above.

\vspace*{3mm}
\section{Conclusions and Discussions}
\vspace*{1.5mm}

 Spontaneous dimensional reduction (SDR) \cite{carlip} is a truly simple and attractive
 concept about the evolution of spacetime in high energies.
 It is well-motivated from many quantum gravity theories, which support
 that the dimension of spacetime is scale-dependent, approaching $\,n=2\,$ in the
 ultraviolet (short distance) and $\,n=4\,$ in the infrared (large distance).
 The world appears four-dimensional to us so far because we have been dealing with
 low energy phenomena at relatively large distances, and the nature of spacetime
 at short distance scales around $\,\mathcal{O}(\text{TeV}^{-1})$\, is still awaiting
 explorations by the LHC.

 In this work, we have studied the exciting possibility that the onset of the SDR happens
 at the TeV scale, under a proper conjecture of dimensional flow (\ref{eq:DFansatz2}).
 We demonstrate that the TeV scale SDR can play a key role
 to restore the unitarity of $WW$ scattering as well as ensuring the renormalizability.
 We have constructed a consistent effective theories of the standard model with SDR
 without or with a Higgs boson (HLSM-SDR and HFSM-SDR),
 in which the electroweak gauge symmetry is nonlinearly realized.
 The model becomes manifestly renormalizable in high energies by simple power counting.
 The model also avoids the fine-tuning problem associated with
 the quadratically divergent radiative corrections to the Higgs mass in the usual 4d SM.

 Then, in Sec.\,3 we analyzed the partial wave unitarity of the $WW$ scattering
 (Fig.\,\ref{Fig-PartWave}) for the HLSM-SDR,
 and computed the longitudinal weak boson scattering cross sections
 in various channels (Figs.\,\ref{fig:cs-wwzz}-\ref{fig:cs-wz-w+w+}).
 Our new mechanism of maintaining unitarity is realized by
 {\it the phase-space reduction of final states under the SDR.}
 These predictions can be tested at the LHC and discriminated from other mechanisms
 of the electroweak symmetry breaking, such as the conventional 4d SM with/without
 a Higgs boson and other ways of unitarization.
 We stress that our new predictions originate from the spontaneous reduction
 of spacetime dimensions at high energies due to the nonperturbative dynamics of
 quantum gravity, so they are universal and show up in all $WW$ scattering channels.
 This is an essential feature of SDR.
 To definitively test the HLSM-SDR and discriminate it from all other EWSB mechanisms requires
 full analysis for all possible $WW$ scattering channels at the second phase of the
 LHC\,(13$-$14\,TeV).\footnote{It is noted that a SM model without the Higgs boson is still
 consistent with the present LHC data, where the newly observed
 125\,GeV boson\,\cite{LHCnew} can be something else, such as a dilaton-like
 scalar \cite{dilaton,Antipin:2013kia}.}

 As shown in Figs.\,\ref{fig:cs-wwzz}-\ref{fig:cs-wz-w+w+},
 our predicted signals display excess above the conventional 4d SM
 (with a relatively light Higgs boson) as a rather broad ``lump" around the
 $0.2-3$\,TeV energy range.
 But as expected, the excess is not as substantial as
 a conventional sharp ``resonance peak"
 like a {600}\,GeV Higgs boson of the SM
 in the $W^+W^-$ channel (Fig.\,\ref{fig:cs-wwzz}).
 According to the experiences with the non-resonance analyses
 at the LHC \cite{nonR,He:2002qi,HVV-2},
 it would be quite challenging for the LHC to measure such broad ``lumps"
 in the $WW$ scattering cross sections, and discriminate them from the
 SM backgrounds.  The second phase of the LHC at 13$-$14\,TeV with a higher integrated
 luminosity around $50-100$\,fb$^{-1}$ or above will be essential for this task.
 In the past, a non-resonance realization of the $WW$ scattering was thought to be
 less exciting among all possible new physics scenarios, but our model provides
 a truly exciting prospect for the non-resonant $WW$ scattering despite the
 experimental challenge,  because probing such non-resonant $WW$ scattering will
 open the door to the {\it quantum spacetimes} with {\it spontaneous dimensional reduction},
 as originated from the dynamics of quantum gravity.

 In Sec.\,4, we studied in parallel the Higgsful SM with the TeV-scale SDR,
 which provides a natural solution to the hierarchy problem \cite{unnatural}
 that troubles the conventional 4d SM.
 For the low energy effective theory with operators up to dimension-4,
 the quantum-gravity-induced new physics effects may be encoded
 in the anomalous gauge couplings of the non-standard Higgs boson,
 as well as the anomalous Higgs self-couplings.
 We found that the non-canceled $E^2$ contributions to the $WW$ scattering
 can be unitarized by the SDR at the TeV scale,
 and the scattering cross sections exhibit different behaviors
 which will be probed at the LHC.
 Fig.\,\ref{fig:uni-dk} summarized the unitarity bounds on the allowed ranges
 of the anomalous gauge coupling $\,\De\ka\,$ of the Higgs boson with mass
 $125$\,GeV.  Unlike the case of the usual 4d SM with $\,\De\ka \neq 0\,$,\,
 these constraints are rather weak on our HFSM-SDR and
 consistent with the current global fits of the LHC data \cite{Hfit}.
 In Fig.\,\ref{fig:cs-wwzz-ac-125}-\ref{fig:cs-wpwp-ac-125},
 we presented the results of a light Higgs boson of mass $\,M_h=125\,$GeV,
 with the anomalous gauge coupling $\,\De\ka \neq 0\,$,\,
 via the scattering processes
 $\,W^+_LW^-_L \to Z_LZ_L$ and $W^\pm_LW^\pm_L \to W^\pm_LW^\pm_L$\,.\,
 Our analysis revealed that
 a non-standard Higgs boson of the HFSM-SDR with mass 125\,GeV has distinctive
 invariant-mass distributions from the naive 4d SM Higgs boson (of the same mass)
 over the $0.2-3$\,TeV energy regions. The signal excesses are manifested via
 broad lumps in {\it all scattering channels} around $\,0.6-3$\,TeV
 energy ranges.  This will be definitively probed by
 the next LHC runs at $13-14$\,TeV collision energies with higher luminosity.

 As the final comment, we compare our SDR constructions
 with the existing extra dimensional models
 under Kaluza-Klein (KK) compactification.
 We note that all SM couplings will become super-renormalizable
 for spacetime dimension $\,n < 4\,$.\,
 In contrast to the KK theories
 which lead to more space dimensions at higher energies, the SDR makes
 spacetime dimensions reduced in the ultraviolet.
 For the KK-type theories, the unitarity of the $S$-matrix for a given longitudinal
 gauge boson scattering is restored due to a geometric Higgs mechanism under which
 the extra components of the higher dimensional gauge fields get eaten by the
 corresponding 4-dimensional KK gauge bosons \cite{HJH-KKuni},
 as characterized by the KK equivalence theorem \cite{HJH-KKuni}.
 In consequence, the $S$-matrix gets unitarized through the exchange of KK states
 which ensure the exact energy cancellations
 and thus the good high energy behavior.
 But including both elastic and inelastic scattering channels in the coupled channel
 analysis still violates the unitarity at a higher scale which must cut off the
 intrinsically nonrenormalizable KK theory \cite{HJH-KKuni}.
 Hence a KK theory could only delay the unitarity violation rather
 than fully removing it \cite{HJH-KKuni}.
 In contrast, for our present HLSM-SDR and HFSM-SDR scenarios, the cross sections
 get unitarized through {\it the universal suppression in phase space.}

 In addition, unlike the usual non-renormalizable 4d Higgsless SM and 5d KK theories,
 our model is renormalizable in the UV and has much better UV behavior due to the SDR.
 Hence it is expected to have better controlled loop-corrections
 and thus better agreement with the precision test than those
 nonrenormalizable theories.
 Studying the electroweak precision constraints
 requires systematical quantum loop calculations under the SDR.
 For this, a method to quantize field theories with varying
 spacetime dimensions is needed, which is fully beyond the present scope and is under
 the current investigations\,\cite{calcagni}\cite{calcagni-2}.
 We will explore this issue for future studies.
 A short summary of the present work is newly given in \cite{He:2013ub}.

\vspace*{3mm}
\begin{appendix}

\section{\,Unitarity Conditions in $n$-Dimensional Spacetime}
\vspace*{1.5mm}

 In this Appendix we systematically derive the unitarity conditions
 for the partial waves and cross sections in general $n$-dimensional spacetime.
 For simplicity and the current analysis, we consider the case in which all
 initial and final particles have the same spin. Extension to more general case
 will be given elsewhere.

  For the cross section in general $n$ dimensions, we need to evaluate the
  corresponding phase space integral of the final states.
  For two-body final states, this integral is given by
 %
 \beqa
   \label{PSInt}
   && \hspace*{-18mm}
   \int\!\di\Pi_2^{(n)}\,\big|{\T(\ECM,\theta)}\big|^2
   \nn
   \\[2mm]
   \hspace*{-10mm}
   &\,=\,&
   \FR{1}{\rh_e^{}}
   \int\!\FR{\di^{n-1}p_1^{}\,\di^{n-1}p_2^{}}
          {(2\pi)^{n-1}(2\pi)^{n-1}}\FR{1}{\,2E_1^{}2E_2^{}\,}
   \big|{\T(\ECM,\theta)}\big|^2(2\pi)^n\de^{(n)}(p-p_1^{}-p_2^{})~~~~~
   \nn
   \\[2mm]
   \hspace*{-10mm}
   &\,=\,&
   \FR{1}{\rh_e^{}}\int\!\FR{\di^{n-1}p_1^{}}{(2\pi)^{n-1}}\FR{1}{\,2E_1^{}2E_2^{}\,}
   \big|{\T(\ECM,\theta)}\big|^2(2\pi)\de(\ECM-E_1^{}-E_2^{}) \,,
 \eeqa
 where $\,\rh_e^{}=1!\,(2!)\,$ is a symmetry factor corresponding to
 the final state particles being nonidentical (identical).
 To complete the remaining integral, we note that the integral measure with
 spherical coordinates in $(n\!-\!1)$-dimensions can be represented by
 \bge
 \label{eq:dp^n-1}
   \di^{n-1}p ~=~
   p^{n-2}\sin^{n-3}\varphi_1^{}\sin^{n-4}\varphi_2^{}\cdots\sin\varphi_{n-3}^{}
   \,\di p\,\di\varphi_1^{}\di\varphi_2^{}\cdots\di\varphi_{n-3}^{}\di\varphi_{n-2}^{} \,.
 \ede
 The region of integration is given by $\,0\leqq p<\infty\,$,\,
 $\,0\leqq \varphi_i^{}\leqq\pi$\, $(1\leqq i\leqq n-3)$\,
 and $\,0\leqq\varphi_{n-2}^{}\leqq 2\pi$\,.\,
 The $\,2\to 2\,$ scattering amplitude depends on c.m.\ energy $\,\ECM$\,
 and a single scattering angle $\,\varphi_1^{}\,$.\,
 Thus, we can always write it in the form $\,f=f\,(p,\varphi_1^{})$\,,\,
 and finish the integration over the rest of angular coordinates
 $\,\varphi_i^{}$ $(1\leqq i\leqq n-3)$\, as follows,
 \beqa
 \int\!\di^{n-1} p\,f(p,\varphi_1^{})
    &\!=\!&
    \int_0^\infty\!\!\!\di p\,p^{n-2}
    \bigg(\prod_{i=2}^{n-3}
    \int_0^{\pi}\!\!\!\di\varphi_i^{}\,\sin^{n-i-2}\varphi_i^{}\bigg)
    \int_0^\pi\!\!\!\di\varphi_1^{}\,\sin^{n-3}\varphi_1^{}
    \int_0^{2\pi}\!\!\!\di\varphi_{n-2}^{}\, f(p,\varphi_1^{})
 \n\\[2mm]
    &\!=\!&
    \int_0^\infty\!\!\!\di p\,p^{n-2}
    \,2\pi\bigg(\prod_{i=2}^{n-3}
    \FR{\sqrt{\pi}\,\Gamma\big(\fr{n-i-1}{2}\big)}{\Gamma(\fr{n-i}{2})}\bigg)
    \int_0^\pi\!\!\!\di\varphi_1^{}\,\sin^{n-3}\varphi_1^{}\, f(p,\varphi_1^{})
 \n\\[2mm]
    &\!=\!&
    \int_0^\infty\!\!\!\di p\,p^{n-2}
    \FR{2\pi^{n/2-1}}{\,\Gamma\big(\fr{n}{2}\!-\!1\big)\,}
    \int_0^\pi\!\!\!\di\varphi_1^{}\,\sin^{n-3}\varphi_1^{}\,
    f(p,\varphi_1^{})   \,.
 \eeqa
 Substituting this result back to the phase space integral (\ref{PSInt}), we derive,
 \beqa
 && \hspace*{-10mm}
 \int\!\!\di\Pi_2^{(n)}\,\big|\T(\ECM,\theta)\big|^2
 \nn\\
 &\,=\,&
 \FR{1}{\,\rh_e(2\pi)^{n-2}\,}\FR{p_1^{n-2}}{\,2E_1 2E_2\,}
 \Big(\FR{p_1}{E_1}\!+\!\FR{p_2}{E_2}\Big)^{\!-1}
 \FR{2\pi^{n/2-1}}{\Gamma(\fr{n}{2}\!-\!1)}
 \int_0^\pi\!\!\!\di\theta\,\sin^{n-3}\theta\big|\T(\ECM,\theta)\big|^2,~~~~~~
 \eeqa
 where we have replaced the remaining angular variable (scattering angle)
 $\,\varphi_1^{}\,$ by the more familiar notation $\,\theta\,$.\,
 In the c.m.\ frame we have,
 \,$E_1^{}=E_2^{}=\ECM/2$\, and $~p_1^{}=p_2^{}\equiv p$\,,\,
 where $\,p_1^{} \equiv |\vec p_1^{}|$\, and \,$p_2^{}\equiv |\vec p_2^{}|$.\,
 Thus, we have,
 \beqa
 \label{PhSpaceInt3}
   \int\!\!\di\Pi_2^{(n)}\,\big|\T(\ECM,\theta)\big|^2
    \,=\, \FR{1}{\,\rh_e^{}2^{n-1}\pi^{n/2-1}\Gamma(\fr{n}{2}\!-\!1)\,}
    \FR{\,p^{n-3}\,}{\ECM}\int_0^\pi\!\!\di\theta\,\sin^{n-3}\theta
    \big|\T(\ECM,\theta)\big|^2 \,,
    \nn\\
 \eeqa
 which holds for $\,n\geqq 3$\,.\,
 For our purpose, this is not quite enough, since we also need to evaluate the integral
 over $\,\di^{n-1} p\,$ with $\,2\leqq n<3\,$.\,
 In the case of $\,2\leqq n<3\,$,\,
 there is no meaning to talk about the angular dependence of the scattering amplitude,
 as commented in Sec.\,3, since the scattering in spacetime dimensions lower than 3
 can be achieved only in the forward or backward direction.
 Thus, we can complete the phase space integral,
 \beq
   \label{PhSpaceInt2}
   \ba{l}
   \dis\int\!\!\di\Pi_2^{(n)}\,
   \left[\big|\T(\ECM,0)\big|^2 + \big|\T(\ECM,\pi )\big|^2\right]
\\[3mm]
   \,=\, \dis
   \FR{\CC}{\,\rh_e^{}2^{n-1}\pi^{(n-3)/2}\Gamma(\fr{n-1}{2})\,}
   \FR{\,p^{n-3}\,}{\ECM}
   \left[\big|\T(\ECM,0)\big|^2 + \big|\T(\ECM,\pi )\big|^2\right] ,
   \hspace*{7mm}
   (\,2\leqq n < 3\,)\,,~~~~
 \ea
 \eeq
 where the coefficient $\,\CC\,$ needs an explanation.
 We note that applying both (\ref{PhSpaceInt3}) and (\ref{PhSpaceInt2}) to $\,n=3\,$
 would yield different results if simply set $\,\CC=1\,$,\,
 due to the improper normalization of phase space in (\ref{PhSpaceInt2}) for
 $\,2\leqq n < 3\,$.\,
 For consistency, we can normalize the coefficient $\,\CC\,$ by imposing the condition
 that the cross section is a smooth function of $\,\ECM$\,.\, So, we have,
 \beqa
   \CC \,=\, \FR{1}{\,\pi\left[|\T(\ECM,0)|^2\!+\!|\T(\ECM,\pi)|^2\right]\,}
   \int_0^\pi\!\!\di\theta\, |\T(\ECM,\theta)|^2 \,.
 \eeqa

  Next, we analyze the unitarity of partial-waves in $n$-dimensions with $\,n>3\,$.\,
  From unitary condition $\,\mathscr{T}^\dag\mathscr{T} = 2\,\IM\mathscr{T}\,$,\,
  we take the matrix element on both sides and insert a complete set of intermediate
  states into its left-hand side. This leads to the condition,
 \beqa
 \label{eq:uni-2N-1}
   \int\!\!\di\Pi_2^{(n)}\big|\T_{\text{el}}(2\!\to\! 2)\big|^2
   +\sum_N\int\!\!\di\Pi_N^{(n)}\,\big|\T_{\text{inel}}(2\!\to\! N)\big|^2
   ~=~ 2\,\IM\T_{\text{el}}(2\!\to\! 2) \,,
 \eeqa
 where $\,\T_{\text{el}}(2\!\to\! 2)$\,
 represents the amplitude for \,$2\to2$\, elastic scattering,
 $\T_{\text{inel}}(2\!\to\! N)$ the amplitudes for \,$2\to N$\,
 ($N\geqq 2$) inelastic scattering,
 and the summation on \,$N$\, runs over all possible intermediate states.
 From (\ref{eq:uni-2N-1}), we have,
 \begin{align}
 \label{PartWaveIneq}
   - \int\!\!\di\Pi_2^{(n)}\,\big|\T_{\text{el}}(2\!\to\! 2)\big|^2
   + 2\IM\T_{\el}(2\!\to\! 2)
   ~=\, \sum_N\int\!\!\di\Pi_N^{(n)}
   \big|\T_{\text{inel}}(2\!\to\! N)\big|^2 \,\geqq\, 0 \,,
 \end{align}
 where the right-hand-side (RHS) is nonnegative as it is a sum of squares.
 Let us expand the elastic amplitude in terms of partial waves for $\,n>3\,$,
 \begin{subequations}
 \label{eq:Tel-al-all}
   \beqa
   \label{eq:Tel-al}
     \T_{\text{el}}(\ECM,\theta) &\,=\,&
     \lam_n^{}\ECM^{4-n}\sum_{\ell}\FR{1}{N_\ell^\nu}
     \C_\ell^\nu(1)\C_\ell^\nu(\cos\theta)\,a^{\text{el}}_\ell(\ECM) \,,
     \\[2mm]
   \label{eq:al-Tel}
     a^{\text{el}}_\ell(\ECM) &\,=\,&
     \FR{\ECM^{n-4}}{\,\lam_n^{}\C_\ell^\nu(1)\,}
     \int_0^\pi\!\!\di\theta\,\sin^{n-3}\theta\,\C_\ell^\nu(\cos\theta)\,
     \T_{\text{el}}(\ECM,\theta) \,,
 \eeqa
 \end{subequations}
 where $\,\lam_n=2(16\pi)^{n/2-1}\Gamma(\fr{n}{2}\!-\!1)$,\, $\nu=\fr{1}{2}(n-3)$,\,  and
 $\,N_\ell^\nu
    =\FR{\pi\Gamma(\ell\!+\! 2\nu)}{\,2^{2\nu\!-\!1}\ell!(\ell+\nu)\Gamma^2(\nu)\,}$.\,
 The function $\,\C_\ell^\nu(x)$\, is the Gegenbauer polynomial of order $\,\nu\,$ and degree
 $\,\ell\,$,\,  and satisfies the following orthogonal condition,
 \beqs
 \beqa
 \label{eq:Cl-Cl'}
 \hspace*{-10mm}&&
 \int_{-1}^1\!\di x\,(1-x^2)^{\nu-\fr{1}{2}}_{}\C_\ell^\nu(x)\C_{\ell'}^\nu(x)
   ~=~ {N_\ell^\nu}\de_{\ell\ell'} \,,
   \\[1.5mm]
 \hspace*{-10mm}&&
  \sum_\ell\FR{1}{N_\ell^\nu}\C_\ell^\nu(x)\C_\ell^\nu(y)
    ~=~ (1-x^2)^{\fr{1-2\nu}{4}}_{}(1-y^2)^{\fr{1-2\nu}{4}}\de(x\!-\!y) \,.
 \eeqa
 \eeqs
 From (\ref{eq:Tel-al-all}), we work out the phase space of the two-body final states,
 \begin{align}
   \label{eq:IntT2}
   \int\!\!\di\Pi_2^{(n)}\big|\T_{\text{el}}(2\to2)\big|^2
   ~=~ \FR{2\lam_n}{\rh_e^{}}\Big(\FR{2p}{\ECM}\Big)^{n-3}\ECM^{4-n}
   \sum_\ell\FR{\big[\C_\ell^\nu(1)\big]^2}{N_\ell^\nu}\big|a_\ell^\el\big|^2 \,.
 \end{align}
 Then we can expand the inequality (\ref{PartWaveIneq}) also in terms of partial waves,
 \beqa
 0 &\,\leqq\,& -\int\di\Pi_2^{(n)}\big|\T_{\text{el}}(2\to2)\big|^2
 + 2\,\IM\T_{\el}(2\to 2)
 \n\\
    &\,=\,&
 \FR{\,2\lam_n^{}\ECM^{4-n}\,}{\rh_e^{}}
 \sum_\ell\FR{\big[\C_\ell^\nu(1)\big]^2}{N_\ell^\nu}
  \bigg[\!-\!\Big(\FR{2p}{\ECM}\Big)^{n-3}|a_\ell^\el |^2
  +\rh_e^{}\FR{\C_\ell^\nu(\cos\theta)}{\C_\ell^\nu(1)}\IM\,a_\ell^\el\bigg] .
  \hspace*{15mm}
 \label{eq:uniExpand}
 \eeqa
 The coefficient outside the brackets on the right-hand side is positive definite.
 Hence, (\ref{eq:uniExpand}) results in,
 \beqa
   0 &\,\leqq\,& -\Big(\FR{2p}{\ECM}\Big)^{n-3}|a_\ell^\el|^2
     +\rh_e^{}\FR{\C_\ell^\nu(\cos\theta)}{\C_\ell^\nu(1)}\IM\,a_\ell^\el \,.
 \eeqa
 At high energies where the masses of initial/final state particles are negligible,
 we have $\,\ECM\simeq 2p\,$.\,  Thus, we deduce,
 \beqa
   0 ~\leqq\, -\big|a_\ell^\el\big|^2+\rh_e^{}\IM\,a_\ell^\el ~=~
   \frac{\,\rh_e^2\,}{4}
   -\big(\RE\,a_\ell^\el\big)^2-\Big(\IM\,a_\ell^\el-\frac{\,\rh_e^{}}{2}\Big)^2
    \,.
 \label{eq:uni-a-el}
 \eeqa
 where we have made use of the property of Gegenbauer polynomial,
 $\,\C_\ell^\nu(1)\geqq |\C_\ell^\nu(x)|$\, for \,$|x|\leqq 1$\,.\,
 From this, we obtain the familiar unitarity condition for each partial wave amplitude,
 \beqa
   \big(\RE\,a_\ell^\el\big)^2
   +\Big(\IM\,a_\ell^\el -\frac{\,\rh_e^{}}{2}\Big)^2 ~\leqq~ \frac{\,\rh_e^2\,}{4} \,.
 \eeqa
 In particular, we have,
 \beqa
 \label{eq:UC-PW-App}
   \big|\RE\,a_\ell^\el \big| ~\leqq~ \FR{\rh_e^{}}{2}\,,
   &~~~~~&
   \big|a_\ell^\el \big| ~\leqq~ \rh_e^{} \,.
 \eeqa
 Similarly, we can derive a unitarity bound for partial wave amplitudes of
 $\,2\to 2\,$ inelastic scattering. This can be done
 by expanding the inelastic amplitude $\,\T_\inel^{}(2\to2)\equiv \T_\inel^{}(\ECM,\theta)$\,
 in terms of partial waves,
 \beqa
   \T_\inel(\ECM,\theta) ~=~
   \lam_n^{} \ECM^{4-n}\sum_{\ell}\FR{1}{N_\ell^\nu}\C_\ell^\nu(1)\C_\ell^\nu(\cos\theta)
   a^\inel_\ell(\ECM) \,.
 \eeqa
 In the high energy region where the masses of initial/final state particles
 are negligible, we infer,
 \begin{align}
 \label{eq:uniExpand-inel}
   \int\!\di\Pi_2^{(n)}\,\big|\T_\inel(2\to2)\big|^2
   ~=~ \FR{\,2\lam_n^{}\ECM^{4-n}\,}{\rh_i^{}}
   \sum_\ell\FR{\big[\C_\ell^\nu(1)\big]^2}{N_\ell^\nu} \big|a_\ell^\inel\big|^2 \,,
 \end{align}
 where $\,\rh_i^{}=1!\,(2!)\,$ is a symmetry factor corresponding to
 the inelastic final state particles being nonidentical (identical).
 Thus, from (\ref{PartWaveIneq}), (\ref{eq:uniExpand})-(\ref{eq:uni-a-el})
 and (\ref{eq:uniExpand-inel}), we have,
 \begin{align}
   &\FR{\,2\lam_n^{}\ECM^{4-n}\,}{\rh_i^{}}\sum_\ell
   \FR{\big[\C_\ell^\nu(1)\big]^2}{N_\ell^\nu}\big|a_\ell^\inel\big|^2
   ~\leqq~ \sum_N\int\!\!\di\Pi_N^{(n)}\big|\T_\inel(2\to N)\big|^2
   \nn \\[2mm]
   &~~= -\int\!\!\di\Pi_2^{(n)}\big|\T_{\text{el}}(2\to2)\big|^2
   +2\IM\T_{\el}(2\to 2)
   \nn \\[2mm]
   &~~\leqq~ \FR{\,2\lam_n^{}\ECM^{4-n}\,}{\rh_e^{}}
   \sum_\ell\FR{\big[\C_\ell^\nu(1)\big]^2}{N_\ell^\nu}\,\FR{\rh_e^2}{4}\,.
 \end{align}
 This leads to the unitarity condition for inelastic partial waves,
 \bge
   \label{UniConInelPWapp}
   \big|a_\ell^\inel\big| ~\leqq~ \FR{\,\sqrt{\rh_i^{}\rh_e^{}}\,}{2} \,.
 \ede

 On the other hand, if we consider that the amplitude is dominated by $s$-wave,
 then we can also derive unitarity bounds for both elastic and inelastic
 scattering cross sections from
 (\ref{PartWaveIneq}), (\ref{eq:IntT2}) and (\ref{eq:UC-PW-App}),
 \beqa
   \si_\el^{}(2\to 2) &\,=\,&
   \FR{1}{2\ECM^2}\int\!\di\Pi_N^{(n)}\big|\T_{\text{el}}(2\to 2)\big|^2
   \nn\\[2mm]
   &=& \FR{\,\lam_n |a_0^\el|^2\,}{\rh_e^{} N_0^\nu}\ECM^{2-n}
   ~\leqq~ \FR{\lam_n \rh_e}{N_0^\nu}\ECM^{2-n} \,,
 \eeqa
 and
 \begin{align}
 \hspace*{-4mm}
   \si_\inel^{}(2\!\to\! N)
    =&~ \FR{1}{\,2\ECM^2\,}\sum_N\!\int\!\!\di\Pi_N^{(n)}\big|\T_{\text{inel}}(2\!\to\! N)\big|^2
  \nn \\[2mm]
    =&~ \FR{1}{\,2\ECM^2\,}
    \bigg[\!-\!\int\!\!\di\Pi_2^{(n)}\big|\T_{\text{el}}(2\!\to\! 2)\big|^2
    +2\IM\T_{el}(2\!\to\!2)\bigg]
    \nn \\[2mm]
    \leqq &~\FR{\,\lam_n^{}\,}{\rh_e}\ECM^{2-n}
    \sum_\ell\FR{\big[\C_\ell^\nu(1)\big]^2}{N_\ell^\nu}
    \bigg[\frac{\,\rh_e^2\,}{4}-\big(\RE\,a_\ell^\el\big)^2
    -\(\IM\,a_\ell^\el-\frac{\rh_e^{}}{2}\)^2
    \bigg]
    \nn \\[2mm]
    \leqq &~ \FR{\lam_n\rh_e}{4N_0^\nu}\ECM^{2-n} \,.
 \end{align}
 Note that both of the above unitarity bounds are smooth functions of spacetime dimension
 \,$n$\,  when \,$n\geqq 2$\, although the original derivation is done with \,$n>3$.\,
 The validity of this extrapolation is guaranteed by the uniqueness of the analytic continuation,
 if we view the unitarity bound as an analytic function of the spacetime dimension $\,n$\,.
 In summary, we have unitarity conditions for both elastic and inelastic cross sections,
 \beqa
 \label{eq:CSuni-n}
 \si_\el^{}(2\!\to\! 2) ~\leqq~ \FR{\lam_n \rh_e}{N_0^\nu}\ECM^{2-n} \,,
 &~~~~~&
 \si_\inel^{}(2\!\to\! N)
 ~\leqq~ \FR{\lam_n\rh_e}{4N_0^\nu}\ECM^{2-n} \,.
 \eeqa
 In 4-dimensions, we have \,$n=4$\, and \,$\nu =\fr{1}{2}\,$.\,
 Thus, these inequalities reduce to
 $\,\si_\el^{}(2\!\to\! 2)\leqq 16\pi\rh_e/\ECM^2\,$ and
 $\,\si_\inel^{}(2\!\to\! N)\leqq 4\pi\rh_e/\ECM^2\,$,\, respectively,
 in accord with the literature.

\end{appendix}

\vspace*{9mm}
\addcontentsline{toc}{section}{Acknowledgments\,}
\noindent
{\bf\large Acknowledgments}
 \\[1mm]
We thank Gianluca Calcagni, Steven Carlip and Dejan Stojkovic
for discussing the spontaneous dimensional reduction, and to Daniel Litim
for discussing the asymptotic safety.
We are grateful to Francesco Sannino and Chris Quigg for discussions
during their visits to Tsinghua HEP Center.
This research was supported by the NSF of China
(under grants 11275101, 10625522, 10635030, 11135003)
and the National Basic Research Program of China (under grant 2010CB833000).


\baselineskip 17pt


\vspace*{4mm}
\begin{thebibliography}{99}
\addcontentsline{toc}{section}{References\,}


\bibitem{SM}
S.\ L.\ Glashow, Nucl.\ Phys.\ {\bf 22} (1961) 579;
S.\ Weinberg, Phys.\ Rev.\ Lett.\ {\bf 19} (1967) 1264;
A.\ Salam, in Elementary Particle Theory, Nobel Symposium No.\,8, edited by
N.\ Svartholm (Almqvist \& Wiksells, Stockholm, 1968), p.\,367.


\bibitem{HM}
P.\ W.\ Higgs, Phys.\ Lett.\ {\bf 12} (1964) 132;
Phys.\ Rev.\ Lett.\ {\bf 13}, 508 (1964); Phys.\ Rev.\ {\bf 145} (1966) 1156;
F.\ Englert and R.\ Brout, Phys.\ Rev.\ Lett.\ {\bf 13} (1964) 321;
G.\ S.\ Guralnik, C.\ R.\ Hagen, and T.\ W.\ Kibble,
Phys.\ Rev.\ Lett.\ {\bf 13} (1964) 585.


\bibitem{tHooft-Veltman}
G.\ 't Hooft, Nucl.\ Phys.\ B\,{\bf 35} (1971) 167;
G.\ 't Hooft and M.\ Veltman,
Nucl.\ Phys.\ B\,{\bf 44} (1972) 189;
Nucl.\ Phys.\ B\,{\bf 50} (1972) 318.


\bibitem{SMunitary}
J.\ M.\ Cornwall, D.\ N.\ Levin, and G.\ Tiktopoulos,
Phys.\ Rev.\ Lett.\ {\bf 30} (1973) 1268;
Phys.\ Rev.\ D\,{\bf 10} (1974) 1145.


\bibitem{SMunitary1}
C.\ H.\ Llewellyn Smith, Phys.\ Lett.\ {\bf 46}B (1973) 233;
D.\ A.\ Dicus and V.\ S.\ Mathur, Phys.\ Rev.\ D\,{\bf 7} (1973) 3111;
B.\ W.\ Lee, C.\ Quigg, and H.\ B.\ Thacker,
Phys.\ Rev.\ Lett.\ {\bf 38} (1977) 883;
Phys.\ Rev.\ D\,{\bf 16} (1977) 1519.


\bibitem{LHCnew}
G.~Aad {\it et al.,}  [ATLAS Collaboration],
Phys.\ Lett.\ B\,{\bf 716} (2012) 1 [arXiv:1207.7214 [hep-ex]].
S.~Chatrchyan {\it et al.,}  [CMS Collaboration],
Phys.\ Lett.\ B\,{\bf 716} (2012) 30 [arXiv:1207.7235 [hep-ex]].


\bibitem{Moriond}
ATLAS and CMS presentations at {\it Rencontres de Moriond 2013} (http://moriond.in2p3.fr),
``EW Interactions and Unified Theories ", March 2-9, 2013, and
``QCD and High Energy Interactions", Mrach 9-16, 2013,
La Thuile, Aosta Valley, Italy.


\bibitem{WW-rev}
M.\ S.\ Chanowitz and M.\ K.\ Gaillard, Nucl.\ Phys.\ B {\bf 261} (1985) 379.
For reviews,  Michael S.\ Chanowitz, arXiv:hep-ph/9812215, and
Czech.\ J.\ Phys.\ {\bf 55} (2005) B45 [arXiv:hep-ph/0412203];
and references therein.


\bibitem{Chanowitz}
M.\ S.\ Chanowitz, M.\ A.\ Furman, I.\ Hinchliffe, Nucl.\ Phys.\ B\,{\bf 153} (1979) 402;
T.\ Appelquist and M.\ S.\ Chanowitz, Phys.\ Rev.\ Lett.\ {\bf 59} (1987) 2405.


\bibitem{Dicus:2004rg}
D.\,A.\ Dicus and H. J. He,
Phys.\ Rev.\ D \textbf{71} (2005) 093009 [hep-ph/0409131];
Phys.\ Rev.\ Lett.\ {\bf 94} (2005) 221802 [hep-ph/0502178].


\bibitem{unnatural}
L.\ Susskind, Phys.\ Rev.\ D\,{\bf 20} (1979) 2619.


\bibitem{trivial}
R.\ Dashen and H.\ Neuberger, Phys.\ Rev.\ Lett.\ {\bf 50} (1983) 1897.


\bibitem{strong}
For a review, C.\ T.\ Hill and E.\ H.\ Simmons,
Phys.\ Rep.\ {\bf 381} (2003) 235 [arXiv:hep-ph/0203079];
and references therein.


\bibitem{susy}
For a recent review,
M.\ E.\ Peskin, arXiv:0801.1928; and references therein.


\bibitem{extrad}
N.\ Arkani-Hamed, S.\ Dimopolous, and G.\ R.\ Dvali,
Phys.\ Lett.\ B\,{\bf 429} (1998) 263;
I.\ Antoniadis, N.\ Arkani-Hamed, S.\ Dimopolous, and G.\ R.\ Dvali,
Phys.\ Lett.\ B\,{\bf 436} (1998) 257;
L.\ Randall and R.\ Sundrum, Phys.\ Rev.\ Lett.\ {\bf 83} (1999) 3370;
Phys.\ Rev.\ Lett.\ {\bf 83} (1999) 4690. 


\bibitem{DC}
N.\ Arkani-Hamed, A.\ G.\ Cohen and  H.\ Georgi,
Phys.\ Rev.\ Lett.\ {\bf 86} (2001) 4757;
C.\ T.\ Hill, S.\ Pokorski and J.\ Wang, Phys.\ Rev.\ D\,{\bf 64} (2001) 105005.


\bibitem{thooft}
G.\ 't Hooft, ``\emph{Dimensional Reduction in Quantum Gravity}",
Salam-festschrift:  \textbf{4}(A) (1993) 1-13 [arXiv:gr-qc/9310026].


\bibitem{carlip}
For reviews, S.\ Carlip,
``{\it The Small Scale Structure of Spacetime}", arXiv:1009.1136 [gr-qc];
``{\it Spontaneous Dimensional Reduction in Short-Distance Quantum Gravity}",
arXiv:0909.3329 [gr-qc];  and references therein.


\bibitem{ERG}
O.\ Lauscher and M.\ Reuter, Phys.\ Rev.\ D\,\textbf{65} (2002) 025013 [arXiv:hep-th/0108040];
Phys.\ Rev.\ D\,{\bf 66} (2002) 025026 [arXiv:hep-th/0205062].


\bibitem{ambjorn}
J.\ Ambj{\o}rn, J.\ Jurkiewicz, and R.\ Loll,
Phys.\ Rev.\ Lett.\ {\bf 95} (2005) 171301 [arXiv:hep-th/0505113];
Phys.\ Rev.\ D\,{\bf 72} (2005)  064014 [arXiv:hep-th/0505154].


\bibitem{LQG}
L. Modesto, 
Class.\ Quant.\ Grav.\ {\bf 26} (2009) 242002 [arXiv:0812.2214 [gr-qc]].


\bibitem{HTstring}
J.\ J.\ Atick and E.\ Witten, Nucl.\ Phys.\ B\,{\bf 310} (1988) 291.


\bibitem{horava}
P.\ Ho\v{r}ava, Phys.\ Rev.\ Lett.\ \textbf{102} (2009) 161301;
Phys.\ Rev.\ D\,\textbf{79} (2009) 084008.


\bibitem{dilaton}
For recent studies,
S.\ Matsuzaki, K. Yamawaki,
Phys.\ Rev.\ D\,{\bf 86} (2012) 035025 [arXiv: 1206.6703];
Phys.\ Rev.\ D\,{\bf 86} (2012) 115004 [arXiv:1209.2017];
Phys.\ Lett.\ B (2013) [arXiv: 1207.5911];
B.\ Bellazzini, C.\ Csaki, J.\ Hubisz, J.\ Serra, and J.\ Terning,
Euro.\ Phys. J.\ C (2013) [arXiv:1209.3299].


\bibitem{Antipin:2013kia}
O.\ Antipin, J.\ Krog, E.\ Molgaard, and F.\ Sannino,
arXiv:1303.7213 [hep-ph]; and references therein.


\bibitem{mureika}
Applications of TeV-scale vanishing dimensions to certain astrophysics and collider phenomenology
were recently considered in a different context,
Jonas R.\ Mureika and Dejan Stojkovic,
Phys.\ Rev.\ Lett.\ \textbf{106} (2011) 101101 [arXiv:1102.3434 [gr-qc]];
Phys.\ Rev.\ Lett.\ {\bf 107} (2011) 169002 [arXiv:1109.3506 [gr-qc]];
L.\ Anchordoqui, D.\ C.\ Dai, M.\ Fairbairn, G.\ Landsberg, and D.\ Stojkovic,
Mod.\ Phys.\ Lett.\ A\,{\bf 27} (2012) 1250021 [arXiv:1003.5914 [hep-ph]];
L.\ A.\ Anchordoqui, D.\ C.\ Dai, H.\ Goldberg, G.\ Landsberg, G.\ Shaughnessy,
D.\ Stojkovic, and T.\ J.\ Weiler, Phys.\ Rev.\ D\,{\bf 83} (2011) 114046
[arXiv:1012.1870 [hep-ph]].


\bibitem{EFT}
For a recent review of the concept of effective field theory,
Steven Weinberg, ``Effective Field Theory, Past and Future",
PoS\,CD\,09 (2009) 001 [arXiv:0908.1964 [hep-th]].


\bibitem{calcagni}
G.\ Calcagni,
Phys.\ Rev.\ Lett.\ \textbf{104} (2010) 251301 [arXiv:0912.3142 [hep-th]];
JHEP {\bf 01} (2012) 065 [arXiv:1107.5041 [hep-th]];
Adv.\ Theor.\ Math.\ Phys.\ {\bf 16} (2012) 549 [arXiv:1106.5787 [hep-th]];
Phys.\ Lett.\ B \textbf{697} (2011) 251 [arXiv:1012.1244 [hep-th]];
Phys.\ Rev.\ D {\bf 84} (2011) 061501 [arXiv:1106.0295 [hep-th]];
JHEP {\bf 1003} (2010) 120 [arXiv:1001.0571 [hep-th]];
M.\ Arzano, G.\ Calcagni, D.\ Oriti, and M.\ Scalisi, Phys.\ Rev.\ D 84 (2011) 125002
[arXiv:1107.5308 [hep-th]];
and references therein.


\bibitem{HVV-0}
R.\ S.\ Chivukula and V.\ Koulovassilopoulos, Phys.\ Lett.\ B\,{\bf 309} (1993) 371
[arXiv:hep-ph/9304293]; V.\ Koulovassilopoulos and R.\ S.\ Chivukula,
Phys.\ Rev.\ D\,{\bf 50} (1994) 3218 [arXiv:hep-ph/9312317].


\bibitem{He:2002qi}
H. J. He, Y.\ P.\ Kuang, C.\ P.\ Yuan and B.\ Zhang, Phys.\ Lett.\ B {\bf 554} (2003) 64
[arXiv:hep-ph/0211229];
and arXiv:hep-ph/0401209, in the proceedings of the
``Workshop on Physics at TeV Colliders", Les Houches, France, May\,26\,--\,June\,6, 2003;
and references therein.


\bibitem{HVV-2}
B.\ Zhang, Y.\ P.\ Kuang, H. J. He, C.\ P.\ Yuan,
Phys.\ Rev.\ D {\bf 67} (2003) 114024 [arXiv:hep-ph/0303048];
T.\ Han, D.\ Krohn, L.\ T.\ Wang and W.\ Zhu, JHEP 1003 (2010) 082 [arXiv: 0911.3656].


\bibitem{DRED}
W.\ Siegel, Phys.\ Lett.\ B\,{\bf 84}, 193 (1979).


\bibitem{DREG}
G. 't Hooft and M. Veltman, Nucl. Phys. B \textbf{44} (1972) 189.


\bibitem{ET}
For a comprehenive review on this subject,
H. J. He, Y.\,P.\ Kuang and C.\,P.\ Yuan,
DESY-97-056 [arXiv:hep-ph/9704276], and references therein.


\bibitem{wein79}
S.\ Weinberg, Physica A\,{\bf 96} (1979) 327.


\bibitem{App}
T.\ Appelquist and C.\ Bernard, Phys.\ Rev.\ D\,{\bf 22} (1980) 200;
A.\ C.\ Langhitano, Phys.\ Rev.\ D\,{\bf 22} (1980) 1166;
Nucl.\ Phys.\ B\,{\bf 188} (1981) 118.


\bibitem{SekharChivukula:2001hz}
 R.\,S.\ Chivukula, D.\,A. Dicus, H. J. He,
 Phys.\ Lett.\ B\,{\bf 525} (2002) 175 [hep-ph/0111016].


\bibitem{schwinger}
J. Schwinger, Phys.\ Rev.\ \textbf{128} (1962) 2425.


\bibitem{deser}
S. Deser, R. Jackiw, and S. Templeton, Ann. Phys. \textbf{140} (1982) 372.


\bibitem{AS-Wein}
S.\ Weinberg, ``Ultraviolet Divergences in Quantum Theories of Gravitation",
in {\it General Relativity: An Einstein Centenary Survey,}
eds.\ S.\ W.\ Hawking and W.\ Israel, Cambridge University Press (1979), p.\,790.


\bibitem{AS-rev}
For recent reviews, D.\ F.\ Litim, ``Renormalisation Group and the Planck Scale",
Phil.\ Trans.\ Roy.\ Soc.\ Lond. A\,{\bf 369} (2011) 2759 [arXiv:1102.4624 [hep-th]];
M.\ Niedermaier and M.\ Reuter, ``The Asymptotic Safety Scenario in Quantum Gravity",
Living Rev.\ Rel.\ {\bf 9} (2006) 5;
and references therein.


\bibitem{soldate}
M. Soldate, Phys. Lett. B \textbf{186} (1987) 321.


\bibitem{K-Pade}
E.g., J.\ L.\ Basdevant, Fortsch.\ Phys.\ {\bf 20} (1972) 283;
S.\ N.\ Gupta, {\it Quantum Electrodynamics}, p.191-198,
Gordon and Breach, New York, 1981;
O.\ Cheyette, M.\ K.\ Gaillard, Phys.\ Lett.\ B\,{\bf 197} (1987) 205;
N.\ Truong, Phys.\ Rev.\ Lett.\ {\bf 61} (1988) 2526;
D.\ Dicus and W.\ Repko, Phys.\ Rev.\ D\,{\bf 42} (1990) 3660;
A.\ Dobado, M.\ J.\ Herrero, and J.\ Terron, Z.\ Phys.\ C {\bf 50} (1991) 205;
and references therein.


\bibitem{Abe:2012fb}
T.\ Abe, N.\ Chen, H. J. He,
JHEP {\bf 1301} (2013) 082 [arXiv:1207.4103];
and references therein.


\bibitem{He:2007ge}
C.\ Du, H. J. He, Y.\ P.\ Kuang, B.\ Zhang, N.\ D. Christensen,
R.\ S.\ Chivukula, and E.\ H.\ Simmons,
Phys.\ Rev.\ D {\bf 86} (2012) 095011 [arXiv:1206.6022];
H. J. He, {\it et al.,} Phys.\ Rev.\ D {\bf 78} (2008) 031701 [arXiv:0708.2588];
A.\,S.\ Belyaev, {\it et al.,}
Phys.\ Rev.\ D {\bf 80} (2009) 055022 [arXiv:0907.2662];
and references therein.



\bibitem{LR}
E.g., R.\ N.\ Mohapatra and G.\ Senjanovic,
Phys.\ Rev.\ D\,{\bf 23} (1981) 165; Phys.\ Rev.\ Lett.\ {\bf 44} (1980) 912.


\bibitem{nonR}
E.g., M.\ S.\ Chanowitz and M.\ K.\ Gaillard, Nucl.\ Phys.\ B\,{\bf 261} (1985) 379;
V.\ D.\ Barger, K.\ Cheung, T.\ Han, R.\,J.\,N.\ Phillips,
Phys.\ Rev.\ D\,{\bf 42} (1990) 3052;
J.\ Bagger {\it et al.,}
Phys.\ Rev.\ D\,{\bf 49} (1994) 1246 [arXiv:hep-ph/9306256];
Phys.\ Rev.\ D\,{\bf 52} (1995) 3878 [arXiv:hep-ph/9504426];
M.\ S.\ Chanowitz and W.\ Kilgore,
Phys.\ Lett.\ B\,{\bf 322} (1994) 147 [arXiv:hep-ph/9311336];
Phys.\ Lett.\ B\,{\bf 347} (1995) 387 [arXiv:hep-ph/9412275];
H. J. He, Y.\ P.\ Kuang and C.\ P. Yuan,
Phys.\ Rev.\ D {\bf 55} (1997) 3038 [arXiv:hep-ph/9611316];
and DESY-97-056 [arXiv:hep-ph/9704276];
M.\ S.\ Chanowitz, Czech.\ J.\ Phys.\ {\bf 55} (2005) B45 [arXiv:hep-ph/0412203];
and references therein.


\bibitem{Sannino}
Roshan Foadi and Francesco Sannino,
Phys.\ Rev.\ D {\bf 78} (2008) 037701 [arXiv:0801.0663];
Roshan Foadi, Matti Jarvinen, and Francesco Sannino,
Phys.\ Rev.\ D {\bf 79} (2009) 035010 [arXiv:0811.3719].


\bibitem{litim}
 P. Fischer and D. F. Litim, Phys. Lett. B \textbf{638} (2006) 497 [arXiv:hep-th/0602203];
 D. F. Litim and T. Plehn, Phys. Rev. Lett. \textbf{100} (2008) 131301 [arXiv:0707.3983];
 J. Hewett and T. Rizzo, JHEP \textbf{0712} (2007) 009 [arXiv:0707.3182];
 E. Gerwick, Eur.\ Phys.\ J.\ C\,\textbf{71} (2011) 1676 [arXiv:1012.1118];
 E. Gerwick, D. Litim and T. Plehn, Phys.\ Rev.\
 D\,\textbf{83} (2011) 084048 [arXiv:1101.5548];
 X.\ Calmet,  Mod.\ Phys.\ Lett.\ A {\bf 26} (2011) 1571 [arXiv:1012.5529].


\bibitem{VacS}
J.\,A.\ Casas, J.\ R.\ Espinosa, M.~Quiros,
Phys.\ Lett.\ B\,{\bf 382} (1996) 374 [arXiv:hep-ph/9603227];
J.\ Ellis, J.\ R.\ Espinosa, G.\ F.\ Giudice, A.\ Hoecker, A.\ Riotto,
Phys.\ Lett.\ B\,{\bf 679}, 369 (2009) [arXiv:0906.0954];
and references therein.


\bibitem{trivial2Loop}
T.\ Hambye and K.\ Riesselmann,
Phys.\ Rev.\ D\,{\bf 55} (1997) 7255 [arXiv:hep-ph/9610272];
and references therein.


\bibitem{EW-rev}
For a recent review, C.\ Quigg, Ann.\ Rev.\ Nucl.\ Part.\ Sci.\ {\bf 59} (2009) 505
[arXiv:0905.3187]; and references therein.


\bibitem{Hfit}
For example, P.\ P.\ Giardino, K.\ Kannike, I.\ Masina, M.\ Raidal, and A.\ Strumia,
arXiv:1303.3570 [hep-ph].


\bibitem{HJH-KKuni}
R.\,S.\ Chivukula, D.\,A.\ Dicus, H. J. He, Phys.\ Lett.\ B \textbf{525} (2002) 175
[hep-ph/0111016];
R.\,S.\ Chivukula and H. J. He,
Phys.\ Lett.\ B {\bf 532} (2002) 121 [hep-ph/0201164];
R.\,S.\ Chivukula, D.\,A.\ Dicus, H. J. He, S.\ Nandi,
Phys.\ Lett.\ B {\bf 562} (2003) 109 [hep-ph/0302263];
H. J. He, Int.\ J.\ Mod.\ Phys.\ A {\bf 20} (2005) 3362 [hep-ph/0412113];
R.\,S.\ Chivukula, H. J. He, M.\ Kurachi, E.\,H.\ Simmons and M.\ Tanabashi,
Phys.\ Rev.\ D {\bf 78} (2008) 095003 [arXiv:0808.1682];
and references therein.


\bibitem{calcagni-2}
Gianluca Calcagni, private communications.


\bibitem{He:2013ub}
 Hong-Jian He and Zhong-Zhi Xianyu,
 Phys. Lett. B \textbf{720} (2013) 142 [arXiv:1301.4570].



\end{thebibliography}
\end{document}